\documentclass[acmsmall]{acmart}
\AtBeginDocument{%
  \providecommand\BibTeX{{%
    \normalfont B\kern-0.5em{\scshape i\kern-0.25em b}\kern-0.8em\TeX}}}

\setcopyright{acmcopyright}
\copyrightyear{2022}
\acmYear{2022}
\acmDOI{nn.nnnn/nnnnnnn.nnnnnnn}

\acmJournal{CSUR}
\acmVolume{1}
\acmNumber{1}
\acmArticle{1}
\acmMonth{1}

\usepackage[longtable]{multirow}
\usepackage{longtable}
\usepackage{multirow}
\usepackage{booktabs}
\usepackage{wasysym}
\usepackage{amsmath,graphicx}
\usepackage{booktabs}
\usepackage{pdflscape}
\usepackage{arydshln}
\usepackage{afterpage}
\usepackage{xcolor}
\begin{document}

\title{Scheduling IoT Applications in Edge and Fog Computing Environments: A Taxonomy and Future Directions}

\author{Mohammad Goudarzi}
%\email{mgoudarzi@student.unimelb.edu.au}
%\orcid{{-5678-9012}
\author{Marimuthu Palaniswami}
%\email{palani@unimelb.edu.au}
\author{Rajkumar Buyya}
%\email{rbuyya@unimelb.edu.au}
\affiliation{%
  \institution{\\The University of Melbourne}
  %\streetaddress{P.O. Box 1212}
  %\city{Melbourne}
  %\state{Cloud Computing and Distributed Systems (CLOUDS) %Laboratory, School of Computing and
  %Information Systems}
  \country{Australia}
  %\email{mgoudarzi@student.unimelb.edu.au}
}

\renewcommand{\shortauthors}{M. Goudarzi et al.}
\begin{abstract}
Fog computing, as a distributed paradigm, offers cloud-like services at the edge of the network with low latency and high-access bandwidth to support a diverse range of IoT application scenarios. To fully utilize the potential of this computing paradigm, scalable, adaptive, and accurate scheduling mechanisms and algorithms are required to efficiently capture the dynamics and requirements of users, IoT applications, environmental properties, and optimization targets. This paper presents a taxonomy of recent literature on scheduling IoT applications in Fog computing. Based on our new classification schemes, current works in the literature are analyzed, research gaps of each category are identified, and respective future directions are described.
\end{abstract}

\begin{CCSXML}
<ccs2012>
<concept>
<concept_id>10002944.10011122.10002949</concept_id>
<concept_desc>General and reference~General literature</concept_desc>
<concept_significance>500</concept_significance>
</concept>
<concept>
<concept_id>10010520.10010521.10010537</concept_id>
<concept_desc>Computer systems organization~Distributed architectures</concept_desc>
<concept_significance>500</concept_significance>
</concept>
</ccs2012>
\end{CCSXML}

\ccsdesc[500]{General and reference~General literature}
\ccsdesc[500]{Computer systems organization~Distributed architectures}
\keywords{Fog computing, Internet of Things, Scheduling Taxonomy, Application Structure, Environmental Architecture, Optimization Characteristics, Performance Evaluation}

\authorsaddresses{Authors' addresses: M. Goudarzi, M. Palaniswami, and R. Buyya, The Cloud Computing and Distributed Systems (CLOUDS) Laboratory, School of Computing and Information Systems, The University of Melbourne, Australia, mgoudarzi@student.unimelb.edu.au}

\maketitle

\section{Introduction}
The Internet of Things (IoT) paradigm has become an integral part of our daily life, thanks to the continuous advancements of hardware and software technologies and ubiquitous access to the Internet. The IoT concept spans a diverse range of application areas such as smart city, industry, transportation, smart home, entertainment, and healthcare, in which context-aware entities (e.g., sensors) can communicate together without any temporal or spatial constraints \cite{gubbi2013internet, hu2017survey}. Thus, it has shaped a new interaction model among different real-world entities, bringing forward new challenges and opportunities. According to Business Insider \cite{businessInsider_2020-md} and International Data Corporation (IDC) \cite{IDC_undated-vl}, 41 Billion active IoT devices will be connected to the Internet by 2027, generating more than 73 Zettabytes of data. The real power of IoT resides in collecting and analyzing the data circulating in the environment \cite{Khodadadi2016}, while the majority of IoT devices are equipped with a constrained battery, computing, storage, and communication units, preventing the efficient execution of IoT applications and data analysis on time. Thus, data should be forwarded to surrogate servers for processing and storage. The processing, storage, and transmission of this gigantic amount of IoT data require special considerations while considering a diverse range of IoT applications.
\subsection{Edge and Fog Computing Paradigms}
Cloud computing is one of the main enablers of IoT that offers on-demand services to process, analyze, and store the data generated from IoT devices in a simplified, scalable, and affordable manner \cite{Khodadadi2016,goudarzi2016mobile, petrakis2018internet}. However, the current cloud infrastructure cannot solely satisfy the requirements of a wide range of IoT applications. First, Cloud Data Centers (CDCs) are located at a multi-hop distance from IoT devices, incurring high access latency and data transmission between IoT devices and CDCs. Thus, it poses an important barrier to the efficient service delivery of real-time and latency-sensitive IoT applications. Besides, the service startup time of IoT applications can be negatively affected due to the extended transmission time required for sending commands to distant Cloud Servers (CSs). Also, the extended transmission period and higher latency lead to higher energy consumption for battery-constrained IoT devices. Second, when myriad IoT devices initiate data-driven interactions with applications deployed on remote CSs, it incurs substantial loads on the network and may lead to severe congestion. Third, it increases the computational overhead on CDCs and may reduce their computing efficiency \cite{madsen2013reliability}. Fourth, the transmission of sensitive raw data over the Internet is not feasible for some IoT applications due to privacy and security concerns \cite{tange2020systematic,meneghello2019iot}. To address these limitations, Edge and Fog computing paradigms have emerged, bringing Cloud-like services to the proximity of IoT devices \cite{bonomi2012fog,Intro-Fog_Computing_principles}.
\par
\begin{figure}[!t]
\centering 
\includegraphics[width=0.6\textwidth, height=5cm]{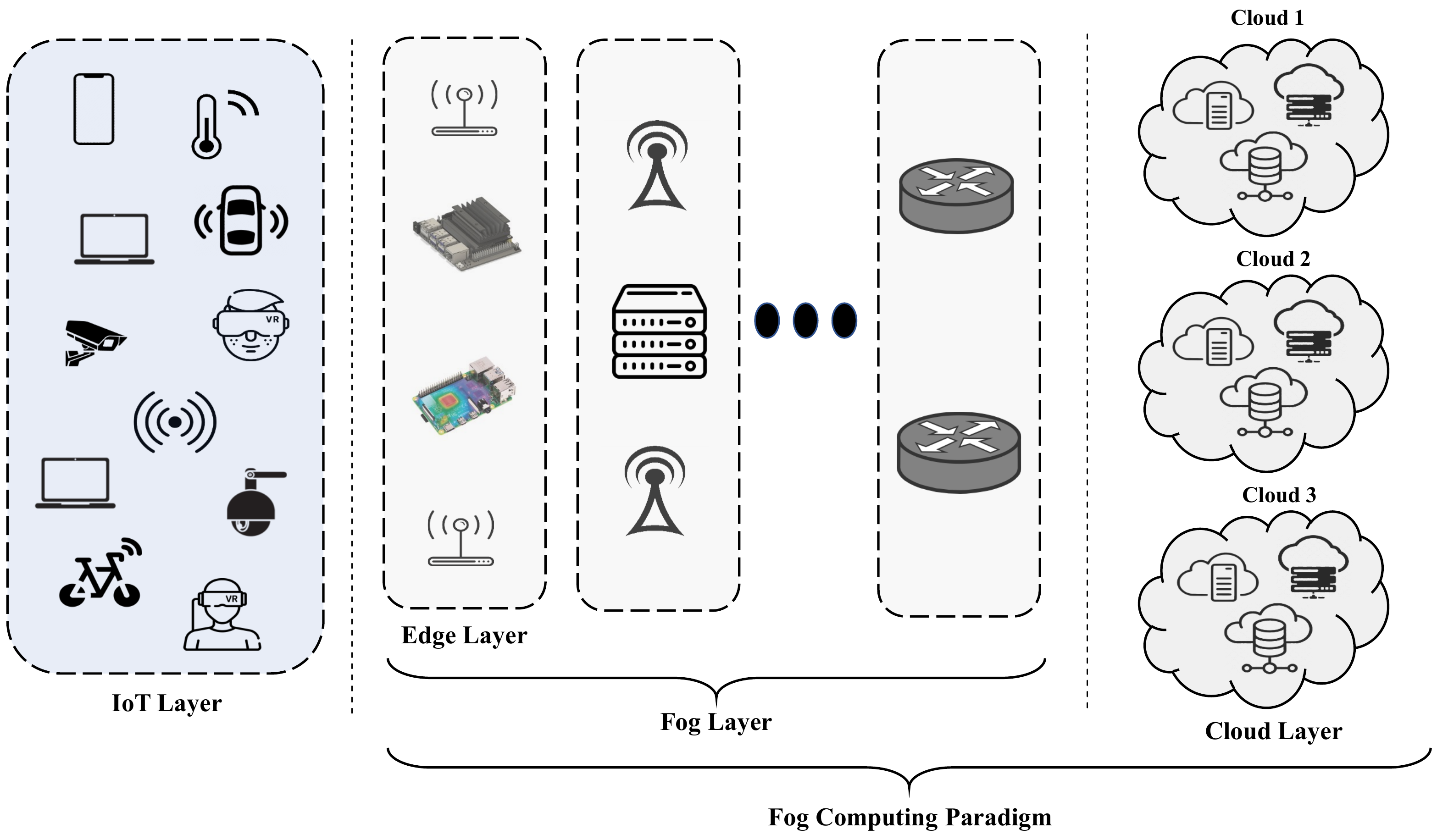}
\caption{An illustration of Fog computing environments}
\label{Fig:basic}
\end{figure}
In the Fog computing paradigm, a large number of geo-distributed and heterogeneous Fog Servers (FSs) are located in an intermediate layer between CSs and IoT devices \cite{yousefpour2019all,goudarzi2019fog}. Distributed FSs (e.g., Raspberry Pis (Rpi), Nvidia Jetson platform, small-cell base stations, nano servers, femtocells, regional servers, core routers, and switches) offer heterogeneous computational and storage resources for IoT devices running various applications, as depicted in Fig.~\ref{Fig:basic}. Since FSs are located in the proximity of IoT devices, compared to CDCs, they can offer Cloud-like services with less latency, which effectively address the requirements of real-time and latency-sensitive IoT applications \cite{yao2018qos}. Moreover, FSs can be accessed with higher bandwidth (i.e., data rate), reducing the required transmission time. Furthermore, Fog computing can help reduce the energy consumption of IoT devices, which is an important parameter, especially for battery-constrained IoT devices. Also, it conserves network bandwidth that decreases the scope of network congestion \cite{mao2017survey,puliafito2019fog}. Besides, Fog computing helps finer computational load distribution, reducing the massive load on CDCs. Finally, Fog computing enables on-premises pre-processing/processing and storage of raw data, which minimizes the requirement of transmitting raw data to distant servers. In this way, we could mitigate the risks of security breaches and preserve data privacy compared to utilizing raw data.
\par
Compared to CSs, FSs usually have limited resources (e.g., CPU, RAM) while they can be accessed more efficiently. Thus, Fog computing does not compete with Cloud computing, but they complement each other to satisfy diverse requirements of heterogeneous IoT applications. In our view, Edge computing harnesses only distributed Edge resources at the closest layer to IoT devices, while Fog computing harnesses distributed resources located in different layers and also Cloud resources (although some works use these terms interchangeably \cite{mach2017mobile,salaht2020overview}), as shown in Fig.~\ref{Fig:basic}.
\subsection{Scheduling IoT Applications in Fog Computing Environments}
Fog computing paradigm provides a scalable solution for integrating a diverse range of hardware and software technologies to offer a wide variety of services for end-users. Fog computing environment is highly heterogeneous in terms of end-users' devices, IoT applications, infrastructures, communication protocols, and deployed frameworks. Hence, the smooth execution of IoT applications in this highly heterogeneous computing environment depends on a large number of contributing parameters, making the efficient scheduling of IoT applications an important and yet a challenging problem in Fog computing environments. To effectively utilize the potential of Fog computing paradigm, these challenges should be thoroughly identified.
\subsubsection{Challenges of Scheduling IoT Applications} The important challenges of scheduling IoT applications are listed below:
\begin{itemize}
    \item \textbf{Challenges related to IoT devices:} The IoT contains a large variety of devices with heterogeneous resource capabilities such as CPU, RAM, storage, and networking \cite{Khodadadi2016}. While IoT devices can be used for the implementation of an unprecedented number of innovative services, their resources are insufficient to host such applications in most scenarios \cite{puliafito2019fog}. Besides, a large category of IoT devices are battery-powered and the direct execution of IoT applications on these devices negatively affects their lifetime. Also, many IoT application scenarios consider moving IoT devices (e.g., real-time patient monitoring systems, drones \cite{mahmud2021ifogsim2}), which further intensify the problem of energy consumption of IoT devices due to increased interference and communication overhead \cite{goudarzi2019fog}.   
   \item \textbf{Challenges related to IoT applications:} The number of IoT applications is rapidly increasing due to recent advancements in technology, emerging business models, and end-users' expectations. The design factors of IoT applications (e.g., granularity level, workload type, and interaction model) heavily depend on application scenario and targeted end-users, necessitating a specific amount of resources for smooth execution and different Quality of Service (QoS) requirements. To illustrate, even in one IoT application domain such as healthcare, there are significantly latency-sensitive applications (e.g., sleep apnea analysis \cite{tuli2019fogbus}) while there are other applications that are more computation-intensive (e.g., radiography \cite{yasser2020covid}). However, the number of computing and communication resources is usually limited. Besides, when several applications are simultaneously requested by the same or different users, the admission control for arriving concurrent requests should be considered as well, making the design factors of an ideal application scheduler even more complex.
    \item \textbf{Challenges related to Edge, Fog, and Cloud resources:} In Fog computing environments, heterogeneous FSs are situated between IoT devices and CDCs through several layers. Usually, it is assumed that FSs are resource-constrained compared to CSs, where FSs at the bottom-most layer (i.e., Edge) have the least amount of resources while providing better access latency and communication bandwidth. Hence, several resource-specific parameters should be considered for scheduling even one IoT application. Besides, considering the ever-increasing number of IoT applications and their diverse resource requirements, some applications cannot be executed on one FS. Thus, FSs require cooperation and resource-sharing with other FSs or CSs for the execution of IoT applications, making the scheduling problem even more complex. However, resource-sharing among FSs is less resilient than CSs due to spatial constraints and high heterogeneity in deployed operating systems, standards, and protocols, just to mention a few \cite{mahmud2020application,madsen2013reliability,lee2019online}. In addition, FSs are more exposed to end-users which makes them potentially less-secured compared to CSs \cite{tange2020systematic}. Besides, the geo-distribution of resources and the IoT data hosted and shared on different servers may also impose privacy implications. As many IoT users may share personal information in Fog computing environments, adversaries can gain access to this shared information \cite{shirazi2017extended}.
     \item \textbf{Challenges related to optimizing parameters} Optimizing the performance of IoT applications running in Fog computing environments depends on numerous parameters such as the main goal of each IoT application, the capabilities of IoT devices, servers properties, networking characteristics, and the imposed constraints. Optimizing the performance of even one IoT application in such a heterogeneous environment with numerous contributing parameters is complex, while multiple IoT applications with different parameters and goals further complicate the problem.
    \item \textbf{Challenges related to decision engines:} Decision engines are responsible to collect all contextual information and schedule IoT applications. Based on the context of IoT applications and environmental parameters, these decision engines may use different optimization modeling \cite{lin2020survey}. Besides, there are several placement techniques to solve these optimization problems. However, considering the types of IoT application scenarios and the number and types of contributing parameters, different placement techniques lead to completely different performance gain \cite{goudarzi2021distributed,tuli2020dynamic}. For example, some placement techniques result in high-accuracy decisions while their decision time takes a long time. However, some other techniques find acceptable solutions with shorter decision time. Moreover, the decision engines can be equipped with several advanced features such as mobility support and failure recovery, enabling them to work in more complex environments.
    \item \textbf{Challenges related to real-world performance evaluation:} The lack of global Fog service providers offering infrastructure on pay-as-you-go models like commercial Cloud platforms such as Microsoft Azure and Amazon Web Services (AWS) pushes researchers to set up small-scale Edge/Fog computing environments \cite{mahmud2021ifogsim2}. The real-world performance evaluation of IoT applications and different placement techniques in Fog computing is not as straightforward as Cloud computing since the management of distributed FSs incurs extra effort and cost, specifically in large-scale scenarios. Besides, the modification and tuning of system parameters during the experiments are time-consuming. Hence, while real-world implementations are the most accurate approach for the performance evaluation of the systems, it is not always feasible, specifically in large-scale scenarios.       
\end{itemize}
\subsubsection{Motivation of Research}
\begin{figure}[!t]
\centering 
\includegraphics[width=0.4\textwidth, height=4cm]{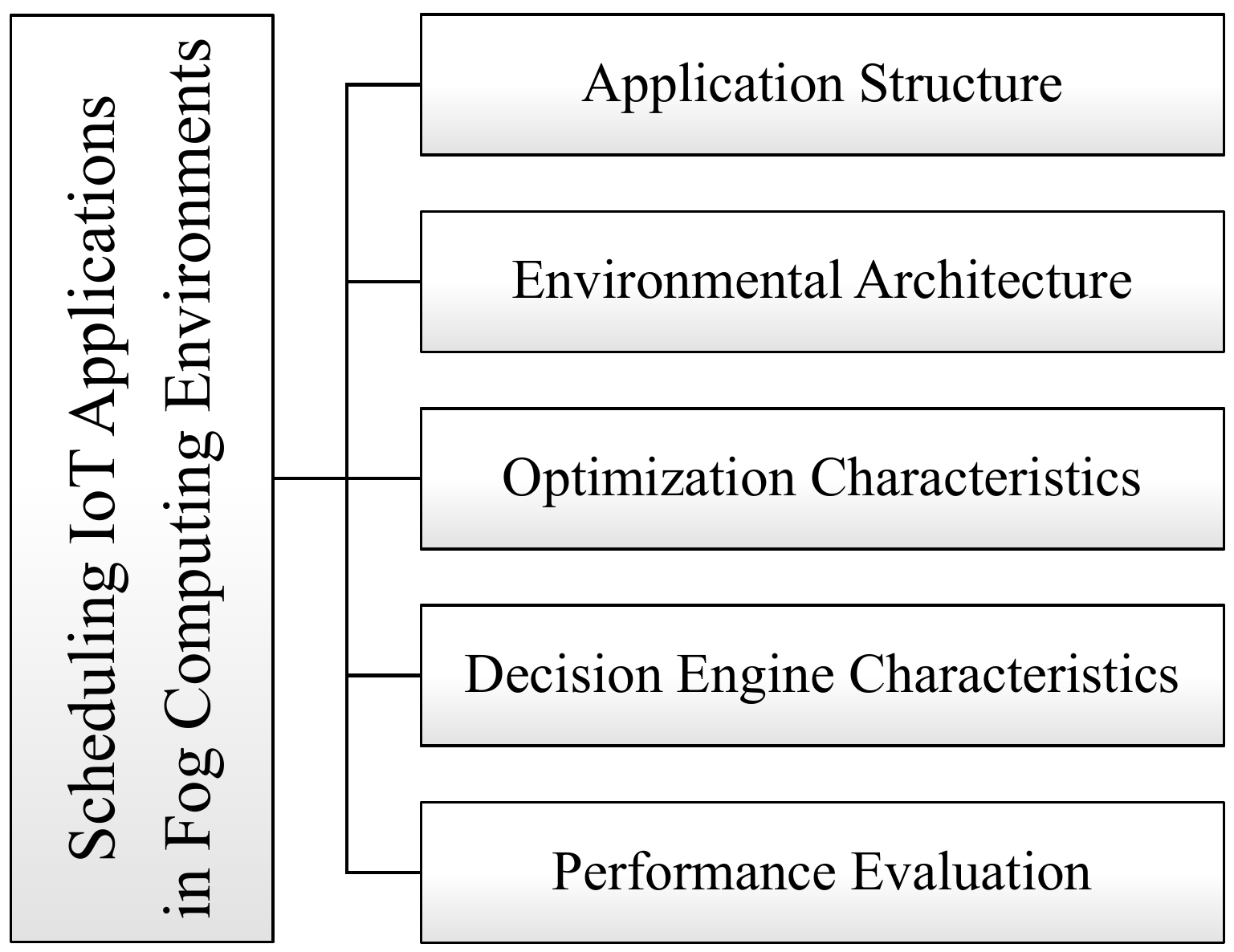}
\caption{Different perspectives of scheduling IoT applications in Fog computing}
\label{Fig:mainCategorization}
\end{figure}

Numerous techniques for scheduling IoT applications in Fog computing environments have been developed to address the above-mentioned challenges. Several works focused on the structure of IoT applications and how these parameters affect the scheduling \cite{mahmud2020application,fiandrino2019profiling,pallewatta2019microservices} while some other techniques mainly focus on environmental parameters of Fog computing, such as the effect of hierarchical Fog layers on the scheduling of IoT applications \cite{karagiannis2020comparison,goudarzi2021distributedmigration}. Besides, several techniques focus on defining specific optimization models to formulate the effect of different parameters such as FSs' computing resources, networking protocols, and IoT devices characteristics, just to mention a few \cite{lin2020survey}. Moreover, several works have proposed different placement techniques to find an acceptable solution for the optimization problem \cite{mahmud2018latency,hu2019learning,abdel2020energy} while some other techniques consider mobility management \cite{dou2020blockchain,verma2021rank,goudarzi2021distributedmigration} and failure recovery \cite{anglano2020profit,goudarzi2020application}. 
All these perspectives directly affect the scheduling problem, especially, when designing decision engines. These perspectives should be simultaneously considered when studying and evaluating each proposal. However, only a few works in the literature have identified the scheduling challenges that directly affect the designing and evaluation of decision engines in Fog computing environments and accordingly categorized proposed works in the literature. Thus, we identify five important perspectives regarding scheduling IoT applications in Fog computing environments, as shown in Fig.~\ref{Fig:mainCategorization}, namely application structure, environmental architecture, optimization modeling, decision engines' characteristics, and performance evaluation.
\par
\begin{figure}[!t]
\centering 
\includegraphics[width=0.6\textwidth, height=4cm]{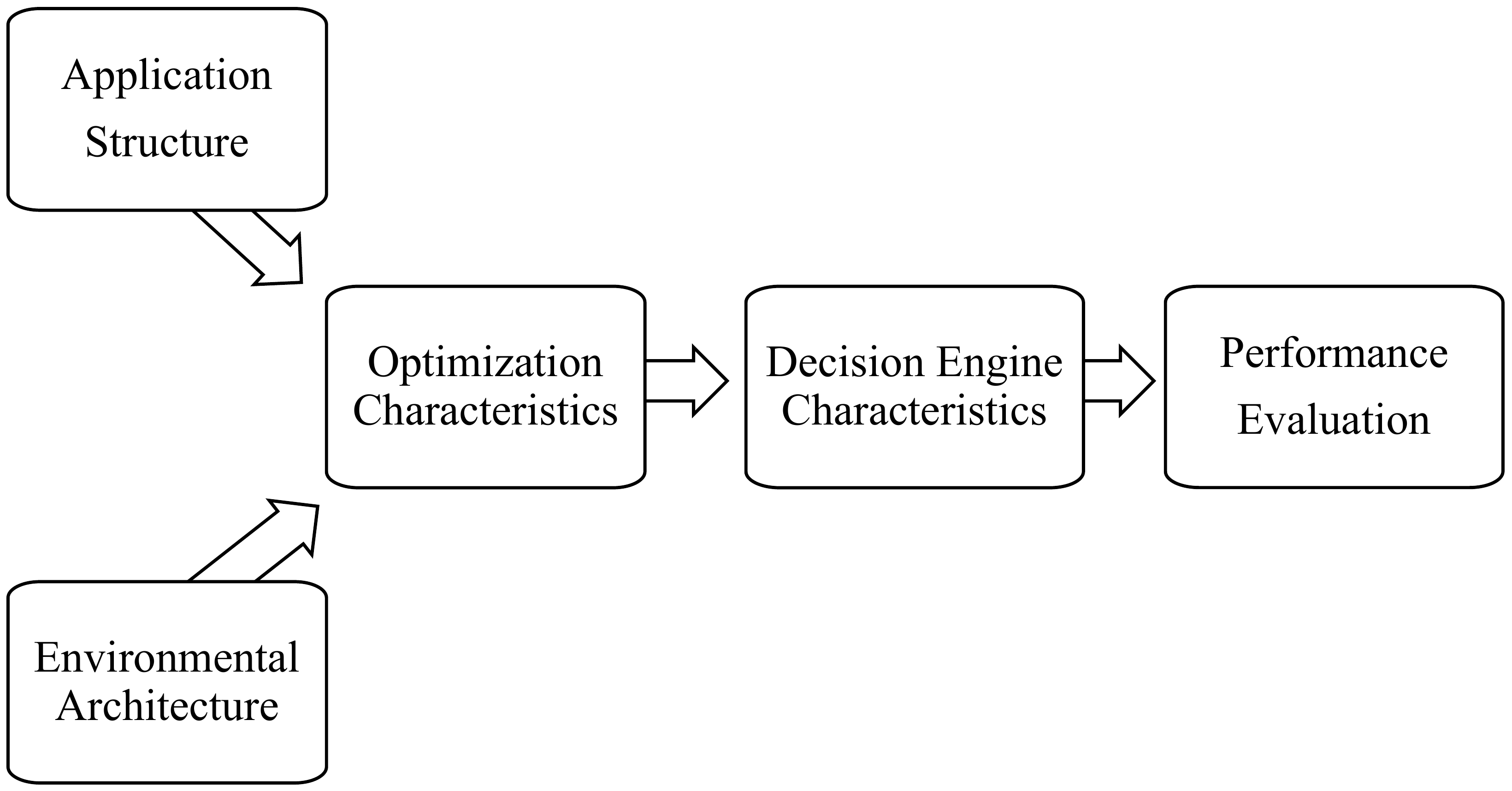}
\caption{The relation among different identified perspectives}
\label{Fig:perspectivesHierarchy}
\end{figure}
Fig.~\ref{Fig:perspectivesHierarchy} depicts the relationships among identified perspectives. The features of application structure and environmental architecture help define the optimization characteristic and formulate the problem. Then, an efficient decision engine is required to effectively solve the optimization problem. Besides, the performance of the decision engine should be monitored and evaluated based on the main goal of optimization for the target applications and environment. Considering each perspective, we present a taxonomy and review the existing proposals. Finally, based on the studied works, we identify the research gaps in each perspective and discuss possible solutions. The main contributions of this work are:
\begin{itemize}
    \item We review the recent literature on scheduling IoT applications in Fog computing from application structure, environmental architecture, optimization modeling, decision engine characteristics, and performance evaluation perspectives and propose separate taxonomies.
    \item We identify research gaps of scheduling IoT applications in Fog computing considering each perspective.
    \item We present several guidelines for designing a scheduling technique in Fog computing paradigm.
    \item We identify and discuss several future directions to help researchers advance Fog computing paradigm.
\end{itemize}
\subsection{Paper Organization}
The rest of the paper is organized as follows. The existing related surveys and taxonomies on scheduling IoT applications in Fog computing environments are studied and compared with ours in Section~\ref{Sec:existingSurveys}. Section~\ref{Sec:applicationStructure} presents a taxonomy and overview of the IoT applications' structure. Section~\ref{Sec:environmentalArchitecture} introduces a taxonomy on environmental properties of resources in Fog computing environments and studies the existing works accordingly. In Section~\ref{Sec:optimizationProblem}, a taxonomy of optimization characteristics of problems in Fog computing environments is introduced. Section~\ref{Sec:decisionEngines} identifies the important aspects of decision engines and presents a taxonomy of decision engines for scheduling IoT applications. Section~\ref{Sec:performanceEvaluation} demonstrates the approaches and metrics used for the evaluation of scheduling strategies in Fog computing. Section~\ref{Sec:designOptions} presents a guideline for designing a scheduling technique. According to identified research gaps, Section~\ref{Sec:futureDirections} provides several future research directions. Finally, Section~\ref{Sec:conclusion} concludes this survey.
\section{Related Surveys}
\label{Sec:existingSurveys}
In the context of Fog computing, surveys targeted different aspects of Fog computing, such as security \cite{shirazi2017extended, roman2018mobile, tange2020systematic, zhang2018security}, smart cities \cite{perera2017fog}, live migration techniques \cite{osanaiye2017cloud}, existing software and hardware \cite{puliafito2019fog}, deep learning applications \cite{wang2020convergence}, healthcare \cite{kraemer2017fog}, and general surveys studied the Fog computing paradigm, its scope, architectures, and recent trends \cite{baccarelli2017fog, mouradian2017comprehensive, hu2017survey, mukherjee2018survey, naha2018fog, nath2018survey, habibi2020fog, singh2021fog}. Also, some surveys mainly discussed resource management, application management, and scheduling in the context of Fog computing, such as \cite{aazam2018offloading, yousefpour2019all, lin2020survey, salaht2020overview, martinez2020design, shakarami2020survey, ghobaei2020resource, mahmud2020application, islam2021context}, that we discuss and compare them with ours.   
\par
Aazam et al.~\cite{aazam2018offloading} reviewed enabling technologies and research opportunities in Fog computing environments alongside studying computation offloading techniques in different domains such as Fog, Cloud, and IoT. Hong et al.~\cite{hong2019resource} and Ghobaei-Arani~\cite{ghobaei2020resource} studied resource management approaches in Fog computing environments and discussed the main challenges for resource management. Yousefpour et al.~\cite{yousefpour2019all} discussed the main features of the Fog computing paradigm and compared it with other related computing paradigms such as Edge and Cloud computing. Besides, it studied the foundations, frameworks, resource management, software, and tools proposed in Fog computing. Mahmud et al.\cite{mahmud2020application} mainly discussed the application management and maintenance in Fog computing and proposed a taxonomy accordingly. Salaht et al.~\cite{salaht2020overview} presented a survey of current research conducted on service placement problems in Fog Computing and categorized these techniques. Shakarami et al.~\cite{shakarami2020survey} studied machine learning-based computation offloading approaches while Adhikari et al. presented the type and applications of nature-inspired algorithms in the Edge computing paradigm. Martinez et al.~\cite{martinez2020design} mainly focused on designing and evaluating Fog computing systems and frameworks. Lin et al.~\cite{lin2020survey} and Sonkoly et al.~\cite{sonkoly2021survey} mainly studied and categorized different approaches for modeling the resources and communication types for computation offloading in Edge computing. Finally, Islam et al.~\cite{islam2021context} proposed a taxonomy for context-aware scheduling in Fog computing and surveyed the related techniques in terms of contextual information such as user and networking characteristics.
\begin{table}[!t]
\centering
\footnotesize
\caption{Related surveys on scheduling in Fog computing}
\resizebox{1\textwidth}{!}{
\renewcommand*{\arraystretch}{1.1}
\begin{tabular}{ccccccccccccc} 
\hline
\multirow{3}{*}{Survey} & \multicolumn{2}{c}{Application Structure}                                           & \multicolumn{2}{c}{Environmental Architecture}                                      & \multicolumn{2}{c}{Optimization Characteristics}                                           & \multicolumn{2}{c}{Decision Engine}                                                 & \multicolumn{2}{c}{Performance Evaluation}                                          & \multirow{2}{*}{\begin{tabular}[c]{@{}c@{}}Conceptualize\\Scheduling\\Framework\end{tabular}} & \multirow{3}{*}{\begin{tabular}[c]{@{}c@{}}Research\\Gap\\(in Years)\end{tabular}}  \\ 
\cmidrule(r){2-3}\cmidrule(r){4-5}\cmidrule(lr){6-6}\cmidrule(lr){7-7}\cmidrule(r){8-9}\cmidrule(r){10-11}
& Taxonomy & \begin{tabular}[c]{@{}c@{}}Research\\Gaps\end{tabular} & Taxonomy  & \begin{tabular}[c]{@{}c@{}}Research\\Gaps\end{tabular} & Taxonomy  & \begin{tabular}[c]{@{}c@{}}Research\\Gaps\end{tabular} & Taxonomy  & \begin{tabular}[c]{@{}c@{}}Research\\Gaps\end{tabular} & Taxonomy  & \begin{tabular}[c]{@{}c@{}}Research\\Gaps\end{tabular} &                                                                                                   &                                                                                    \\ 
\cmidrule(lr){1-1}\cmidrule(r){2-3}\cmidrule(r){4-5}\cmidrule(r){6-7}\cmidrule(lr){8-9}\cmidrule(lr){10-11}\cmidrule(lr){12-12}\cmidrule(lr){13-13}
\cite{aazam2018offloading}                  & \scalebox{1.5}{\ensuremath \Circle}        & \scalebox{1.5}{\ensuremath \LEFTcircle}                                                                & \scalebox{1.5}{\ensuremath \Circle}        & \scalebox{1.5}{\ensuremath \Circle}                                                                     & \scalebox{1.5}{\ensuremath \Circle}        & \scalebox{1.5}{\ensuremath \Circle}                                                                      & \scalebox{1.5}{\ensuremath \Circle}        & \scalebox{1.5}{\ensuremath \Circle}                                                                      & \scalebox{1.5}{\ensuremath \Circle}        & \scalebox{1.5}{\ensuremath \Circle}                                                                      & \scalebox{1.5}{\ensuremath \CIRCLE}                                                                                             & 4                                                                                  \\
\cite{hong2019resource}      & \scalebox{1.5}{\ensuremath \Circle}       &  \scalebox{1.5}{\ensuremath \Circle}                                                                      & \scalebox{1.5}{\ensuremath \LEFTcircle}  & \scalebox{1.5}{\ensuremath \LEFTcircle}                                                               & \scalebox{1.5}{\ensuremath \LEFTcircle}  & \scalebox{1.5}{\ensuremath \Circle}                                                                      & \scalebox{1.5}{\ensuremath \LEFTcircle}  & \scalebox{1.5}{\ensuremath \Circle}                                                                      & \scalebox{1.5}{\ensuremath \Circle}        & \scalebox{1.5}{\ensuremath \Circle}                                                                      & \scalebox{1.5}{\ensuremath \Circle}                                                                                                & 3.5                                                                                \\
\cite{yousefpour2019all}          & \scalebox{1.5}{\ensuremath \Circle}      & \scalebox{1.5}{\ensuremath \Circle}                                                                       & \scalebox{1.5}{\ensuremath \LEFTcircle}  & \scalebox{1.5}{\ensuremath \LEFTcircle}                                                                & \scalebox{1.5}{\ensuremath \Circle}        & \scalebox{1.5}{\ensuremath \Circle}                                                                      & \scalebox{1.5}{\ensuremath \Circle}        & \scalebox{1.5}{\ensuremath \LEFTcircle}                                                                & \scalebox{1.5}{\ensuremath \Circle}       & \scalebox{1.5}{\ensuremath \Circle}                                                                     & \scalebox{1.5}{\ensuremath \Circle}                                                                                                & 3                                                                                  \\
\cite{ghobaei2020resource}       & \scalebox{1.5}{\ensuremath \Circle}       &   \scalebox{1.5}{\ensuremath \Circle}                                                                     & \scalebox{1.5}{\ensuremath \Circle}        & \scalebox{1.5}{\ensuremath \Circle}                                                                      & \scalebox{1.5}{\ensuremath \Circle}        & \scalebox{1.5}{\ensuremath \Circle}                                                                      & \scalebox{1.5}{\ensuremath \LEFTcircle}  & \scalebox{1.5}{\ensuremath \LEFTcircle}                                                                & \scalebox{1.5}{\ensuremath \Circle}        & \scalebox{1.5}{\ensuremath \Circle}                                                                      & \scalebox{1.5}{\ensuremath \Circle}                                                                                                & 2.5                                                                                \\
\cite{salaht2020overview}              & \scalebox{1.5}{\ensuremath \Circle}       & \scalebox{1.5}{\ensuremath \Circle}                                                                       & \scalebox{1.5}{\ensuremath \Circle}        & \scalebox{1.5}{\ensuremath \Circle}                                                                     & \scalebox{1.5}{\ensuremath \LEFTcircle}  & \scalebox{1.5}{\ensuremath \CIRCLE}                                                                      & \scalebox{1.5}{\ensuremath \LEFTcircle}  & \scalebox{1.5}{\ensuremath \LEFTcircle}                                                                & \scalebox{1.5}{\ensuremath \CIRCLE}       & \scalebox{1.5}{\ensuremath \LEFTcircle}                                                                & \scalebox{1.5}{\ensuremath \Circle}                                                                                                & 2                                                                                  \\
\cite{mahmud2020application}             & \scalebox{1.5}{\ensuremath \CIRCLE}       & \scalebox{1.5}{\ensuremath \CIRCLE}                                                                       & \scalebox{1.5}{\ensuremath \Circle}        & \scalebox{1.5}{\ensuremath \LEFTcircle}                                                                & \scalebox{1.5}{\ensuremath \Circle}        & \scalebox{1.5}{\ensuremath \Circle}                                                                     & \scalebox{1.5}{\ensuremath \LEFTcircle}  & \scalebox{1.5}{\ensuremath \LEFTcircle}                                                                & \scalebox{1.5}{\ensuremath \Circle}        & \scalebox{1.5}{\ensuremath \Circle}                                                                      & \scalebox{1.5}{\ensuremath \CIRCLE}                                                                                                & 1.5                                                                                \\
\cite{shakarami2020survey}           & \scalebox{1.5}{\ensuremath \Circle}       & \scalebox{1.5}{\ensuremath \Circle}                                                                       & \scalebox{1.5}{\ensuremath \Circle}        & \scalebox{1.5}{\ensuremath \Circle}                                                                      & \scalebox{1.5}{\ensuremath \Circle}        & \scalebox{1.5}{\ensuremath \Circle}                                                                      & \scalebox{1.5}{\ensuremath \LEFTcircle}  & \scalebox{1.5}{\ensuremath \LEFTcircle}                                                                & \scalebox{1.5}{\ensuremath \Circle}       & \scalebox{1.5}{\ensuremath \Circle}                                                                     & \scalebox{1.5}{\ensuremath \Circle}                                                                                                & 1.5                                                                                \\
\cite{martinez2020design}            & \scalebox{1.5}{\ensuremath \Circle}       & \scalebox{1.5}{\ensuremath \Circle}                                                                       & \scalebox{1.5}{\ensuremath \Circle}        & \scalebox{1.5}{\ensuremath \Circle}                                                                     & \scalebox{1.5}{\ensuremath \Circle}        & \scalebox{1.5}{\ensuremath \Circle}                                                                      & \scalebox{1.5}{\ensuremath \Circle}        & \scalebox{1.5}{\ensuremath \LEFTcircle}                                                                & \scalebox{1.5}{\ensuremath \LEFTcircle}  & \scalebox{1.5}{\ensuremath \LEFTcircle}                                                                & \scalebox{1.5}{\ensuremath \Circle}                                                                                                & 1.5                                                                                \\
\cite{lin2020survey}          &   \scalebox{1.5}{\ensuremath \Circle}     & \scalebox{1.5}{\ensuremath \Circle}                                                                       & \scalebox{1.5}{\ensuremath \LEFTcircle}  & \scalebox{1.5}{\ensuremath \Circle}                                                                     & \scalebox{1.5}{\ensuremath \CIRCLE}        & \scalebox{1.5}{\ensuremath \LEFTcircle}                                                                & \scalebox{1.5}{\ensuremath \Circle}        & \scalebox{1.5}{\ensuremath \LEFTcircle}                                                                & \scalebox{1.5}{\ensuremath \Circle}        & \scalebox{1.5}{\ensuremath \Circle}                                                                      & \scalebox{1.5}{\ensuremath \Circle}                                                                                               & 1.5                                                                                \\
\cite{islam2021context}               & \scalebox{1.5}{\ensuremath \Circle}       & \scalebox{1.5}{\ensuremath \LEFTcircle}                                                                 & \scalebox{1.5}{\ensuremath \LEFTcircle}  & \scalebox{1.5}{\ensuremath \Circle}                                                                      & \scalebox{1.5}{\ensuremath \Circle}        & \scalebox{1.5}{\ensuremath \Circle}                                                                      & \scalebox{1.5}{\ensuremath \Circle}        & \scalebox{1.5}{\ensuremath \LEFTcircle}                                                                & \scalebox{1.5}{\ensuremath \Circle}        & \scalebox{1.5}{\ensuremath \Circle}                                                                      & \scalebox{1.5}{\ensuremath \Circle}                                                                                               & 1          \\
\cite{sonkoly2021survey} &\scalebox{1.5}{\ensuremath \LEFTcircle}  & \scalebox{1.5}{\ensuremath \Circle}&\scalebox{1.5}{\ensuremath \Circle} &\scalebox{1.5}{\ensuremath \Circle} &\scalebox{1.5}{\ensuremath \LEFTcircle} &\scalebox{1.5}{\ensuremath \LEFTcircle}  &\scalebox{1.5}{\ensuremath \Circle}  &\scalebox{1.5}{\ensuremath \Circle}  &\scalebox{1.5}{\ensuremath \Circle}  &\scalebox{1.5}{\ensuremath \Circle}  & \scalebox{1.5}{\ensuremath \Circle}  & 0.5
\\
\cite{adhikari2021comprehensive}            & \scalebox{1.5}{\ensuremath \Circle}       & \scalebox{1.5}{\ensuremath \Circle}                                                                       & \scalebox{1.5}{\ensuremath \Circle}        & \scalebox{1.5}{\ensuremath \Circle}                                                                      & \scalebox{1.5}{\ensuremath \Circle}        & \scalebox{1.5}{\ensuremath \LEFTcircle}                                                                & \scalebox{1.5}{\ensuremath \Circle}        & \scalebox{1.5}{\ensuremath \LEFTcircle}                                                                & \scalebox{1.5}{\ensuremath \Circle}        & \scalebox{1.5}{\ensuremath \Circle}                                                                      & \scalebox{1.5}{\ensuremath \Circle}                                                                                                & 0.5                                                                                \\
This Survey             & \scalebox{1.5}{\ensuremath \CIRCLE}       & \scalebox{1.5}{\ensuremath \CIRCLE}                                                                       & \scalebox{1.5}{\ensuremath \CIRCLE}        & \scalebox{1.5}{\ensuremath \CIRCLE}                                                                      & \scalebox{1.5}{\ensuremath \CIRCLE}       & \scalebox{1.5}{\ensuremath \CIRCLE}                                                                      & \scalebox{1.5}{\ensuremath \CIRCLE}        & \scalebox{1.5}{\ensuremath \CIRCLE}                                                                      & \scalebox{1.5}{\ensuremath \CIRCLE}        & \scalebox{1.5}{\ensuremath \CIRCLE}                                                                      & \scalebox{1.5}{\ensuremath \CIRCLE}                                                                                                & Current                                                                            \\
\hline
\multicolumn{13}{l}{
\scalebox{1.5}{\ensuremath \CIRCLE}: Full Cover, \scalebox{1.5}{\ensuremath \LEFTcircle}: Partial Cover, \scalebox{1.5}{\ensuremath \Circle} : Does Not Cover
}
\end{tabular}
}
\label{Tab:relatedSurveys}
\end{table}
\par
Table~\ref{Tab:relatedSurveys} summarizes the characteristics of related surveys and provides a qualitative comparison with our work. The proper scheduling of IoT applications in Fog computing environments can be viewed from different perspectives, such as application structure, environmental architecture, optimization modeling, and the features of decision engines. Besides, the performance of scheduling techniques should be continuously evaluated to offer the best performance. As depicted in Table~\ref{Tab:relatedSurveys}, the existing surveys barely study and provide comprehensive taxonomy for the above-mentioned perspectives. In this work, we identify the main parameters of each perspective and present a taxonomy accordingly. Moreover, we identify related research gaps and provide future directions to improve the Fog computing paradigm.
\section{Application Structure}
\label{Sec:applicationStructure}
\begin{figure}[!t]
\centering 
\includegraphics[width=0.5\textwidth, height=5cm]{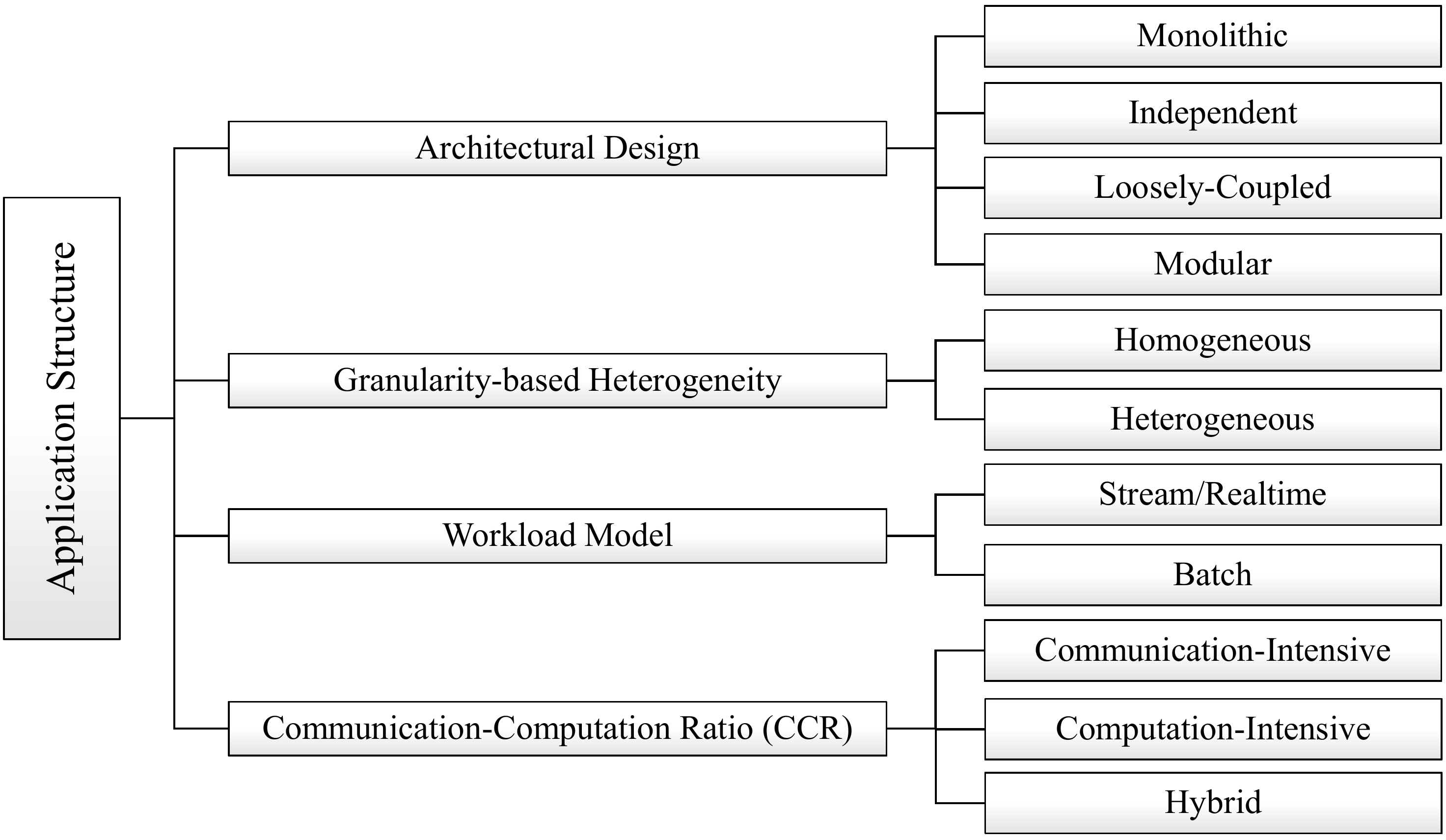}
\caption{Application structure taxonomy}
\label{Fig:applicationStructureTaxonomy}
\end{figure}

The primary goal of Fog computing is to offer efficient and high-quality service to users with heterogeneous applications and requirements. Hence, service providers require a comprehensive understanding of each IoT application structure (e.g., workload model and latency requirements) to better capture its complexities, perform efficient scheduling and resource management, and offer high-quality service to the users. Also, when designing the architecture of each IoT application, dynamics, constraints, and complexities of resources in Fog computing should be carefully considered to exploit the potential of this paradigm. Fig.~\ref{Fig:applicationStructureTaxonomy} presents a taxonomy and main elements of IoT application structure, described below.
\subsection{Architectural Design}
The logic of IoT applications can be implemented in different ways. To illustrate, operations of a Video Optical Character Recognition (VideoOCR) such as capturing frames, similarity check, and text extraction can be implemented as a single encapsulated program or as a set of interdependent components \cite{deng2021fogbus2}. Hence, according to the granularity level of applications, their distribution, and coupling intensity, the architectural design of applications can be classified into four types:
\subsubsection{Monolithic} It encapsulates the complete logic of an application as a single component or program. The parallel execution of these applications can be obtained using multi processing approaches \cite{mahmud2020application}. In the context of Fog computing, several works such as \cite{bahreini2021vecman,min2019learning,maleki2021mobility,chen2018optimized} have considered monolithic applications. 
\subsubsection{Independent} These applications require the execution of a set of independent tasks or components for the complete execution of the application. The constituent parts of these applications can be simultaneously executed on different FSs or CSs. Several works such as \cite{asheralieva2019learning,huang2018distributed,hazra2021ceco,peng2021constrained} discuss applications with independent components or tasks in the Fog literature.
\subsubsection{Modular} Each modular application is composed of a set of dependent tasks or components. While constituent parts of each application can be distributed over several FSs or CSs for parallel execution, there are some constraints for the execution of tasks based on their data-flow dependency model. Several works in the literature such as \cite{goudarzi2021distributed,liu2021efficient,lu2020optimization,peixoto2021hierarchical} discuss modular applications. 
\subsubsection{Loosely-coupled} Components of loosely-coupled applications (i.e., microservices) can be distributed over several CSs or FSs. Besides, due to service-level isolation, components of applications can be shared among different applications, providing high application extendability. Several works such as \cite{deng2020optimal,wang2019delay,goudarzi2021distributedmigration,deng2021fogbus2} have considered loosely-coupled applications.  
\subsection{Granularity-based Heterogeneity}
Tasks within an IoT application have different properties such as computation size, input size, output size, and deadline. These features affect the scheduling complexity, where identifying the dynamics of applications with heterogeneous task properties requires further effort. Accordingly, we categorize IoT applications based on their granularity-level specifications to 1) \textbf{heterogeneous} such as \cite{liu2021efficient,ho2020joint} or 2) \textbf{homogeneous} such as \cite{huang2018distributed,zhang2019task}.     
\subsection{Workload Model}
The workload model specifies the data architecture of an application, which can be broadly divided into two categories for IoT applications:
\subsubsection{Stream/Realtime} In this category, the data should be processed by the servers as soon as it was generated (i.e., real-time), and hence, the data usually require relatively simple transformation or computation. Several works such as \cite{cheng2021multi,kishor2021task,maia2021improved} discuss stream workload for IoT applications.   
\subsubsection{Batch} In batch processing, the input data of an application is usually bundled for processing. However, contrary to heavy batch processing models, IoT applications often use micro-batches to provide a near-realtime experience. In the literature, several works such as \cite{wang2019dynamic,han2019ondisc,chen2018multi} consider batch workload for the applications.
\subsection{Communication-Computation Ratio (CCR)} Each IoT application, regardless of its architecture, contains some amount of input data for transmission and computational load for processing. These properties can significantly affect the scheduling decision to find proper FSs or CSs for an application. The CCR defines whether an application on average is more 1) \textbf{computation-intensive} \cite{wang2020joint,guo2018efficient,pallewatta2019microservices} or 2) \textbf{communication-intensive} \cite{hong2019multi,sarkar2021collaborative,bahreini2021vecman}. Besides, some works consider a range of applications to cover both computation-intensive and communication-intensive applications, to which we refer as 3) \textbf{hybrid}  \cite{cai2021failure,goudarzi2020application}.
\subsection{Discussion}
In this section, we discuss the effects of identified application structure's elements on the decision engine and describe the lessons that we have learned. Besides, we identify several research gaps accordingly. Table~\ref{Tab:ApplicationStructure} summarizes the characteristics related to IoT application structure in Fog computing.
\subsubsection{\textbf{Effects on the decision engine}} The application structure properties affect the decision engine in various aspects, as briefly described below. 
\paragraph{1} Architectural design: It defines the number of tasks/modules and their respective dependency within a single application. Hence, as the number of tasks/modules per application increases, the problem space significantly increases. Considering an application with $n$ number of tasks and $m$ possible candidate configuration per task, the Time Complexity (TC) of finding the optimal solution is $O(m^n)$. Besides, the dependency of tasks within an application imposes hard constraints on the problem, which further increases the complexity. Thus, finding the optimal solution for the scheduling of applications becomes very time-consuming, and the design of an efficient placement technique to serve applications in a timely manner remain an important yet challenging problem.  
\paragraph{2} Granularity-based heterogeneity: It shows the corresponding properties of each task/module within an application and plays a principal role in identifying the dynamics of applications. One of the most important features of decision engines is their adaptability and their capability to extract the complex dynamics of applications so that the decision engine can receive diverse types of applications' requests. Since applications with heterogeneous granularity-based properties have higher dynamics' complexity, the decision engines designed for this application category should support high adaptability. 
\paragraph{3} Workload model and CCR: These elements provide insightful information regarding the input data architecture of the application and its behavior in the runtime. Accordingly, the decision engine may define different priority queues for incoming requests based on their workload model and CCR to provide higher QoS for the users. For example, applications with real-time workload types and communication-intensive CCR may have higher priority for the placement on servers closer to the IoT devices than computation-intensive applications that are not real-time.
\subsubsection{\textbf{Lessons learned}} Our findings regarding the IoT application structure in the surveyed works are briefly described in what follows:
\paragraph{1} Almost 70\% of the surveyed works have overlooked studying the dependency model of tasks within an application and selected either the independent or monolithic design. The rest of the works consider dependency among tasks of an application in different models (i.e., sequential, parallel, or hybrid dependency). Moreover, only about 10\% of the studied works consider microservices in their application design.   
\paragraph{2} The most realistic assumption for the granular properties of each task/module is heterogeneous (i.e., heterogeneous input size, output size, and computation size). Almost 85\% of the studied works consider heterogeneous properties for each task/module, while around 15\% of the works consider the homogeneous properties for the tasks/modules. 
\paragraph{3} The workload model and CCR in each proposal depend on the targeted application scenarios. Almost 55\% of the works did not study the CCR, or the required information to obtain the CCR (i.e., computation size of tasks, average data size) was not mentioned. Among the rest of the works, computation-intensive, communication-intensive, and hybrid CCR form roughly 25\%, 15\%, and 5\% of proposals respectively. 

\subsubsection{\textbf{Research Gaps}}
We have identified several open issues for further investigation, that are discussed below:   
\paragraph{1} According to Alibaba’s data of 4 million applications, more than 75\% of the applications consist of dependent tasks \cite{zhao2021offloading}. However, only around 30\% of the recent works surveyed in this study consider applications with dependent tasks (i.e., modular or loosely-coupled), showing further investigation is required to identify the dynamics of these types of complex applications.
\paragraph{2} Although the microservice-based applications can significantly benefit the IoT scenarios, only a few works such as \cite{wang2019delay,goudarzi2021distributedmigration} have studied the scheduling and migration of microservices in Edge/Fog computing environments. So, further investigation is required to study the behavior of microservice-based applications with different resource management techniques.
\paragraph{3} Modular or loosely-coupled IoT applications can be distributed over different FSs or CSs for parallel execution. However, several works such as \cite{mahmud2018latency} statically assign components of an application on pre-defined FSs or CSs and only schedule one or two remaining components. Hence, the best placement configuration of applications based on the current dynamics of the system cannot be investigated, leading to diminished performance gain.
\paragraph{4} When the number of IoT applications increases, there is a high probability that application requests with different workload models are submitted to the system. However, none of the studied works in the literature consider applications with different workload models and how they may mutually affect each other in terms of performance.
\paragraph{5} Due to the high heterogeneity of IoT applications in Fog, applications with diverse CCR may be submitted for processing, requiring special consideration such as networking and prioritization. Although there are only a few recent works such as \cite{goudarzi2021distributed,cheng2021multi} that consider hybrid CCR, most of the recent works target one of the computation-intensive or communication-intensive applications.
\scriptsize
\begin{longtable}[H]{ccccc||ccccc}
\caption{Summary of existing works considering application structure taxonomy \label{Tab:ApplicationStructure}}\\
\hline
\multirow{2}{*}{Ref}                    & \multicolumn{4}{c||}{Application Structure}                                                                                                      & \multirow{2}{*}{Ref}                             & \multicolumn{4}{c}{Application Structure}                                                                                                         \\* 
\cline{2-5}\cline{7-10}
& Design          & \begin{tabular}[c]{@{}c@{}}Granular\\Hetero\end{tabular} & \begin{tabular}[c]{@{}c@{}}Workload\\Model\end{tabular} & CCR    &                                                  & Design          & \begin{tabular}[c]{@{}c@{}}Granular\\Hetero\end{tabular} & \begin{tabular}[c]{@{}c@{}}Workload\\Model\end{tabular} & CCR     \endfirsthead 
\hline
\cite{asheralieva2019learning} & Independent     & Hetero                                                   & Batch                                                   & Comp   & \cite{mahmud2019context}                & Modular         & Hetero                                                   & Stream                                                  & ND      \\ 
\hdashline
\cite{bahreini2021vecman}      & Monolithic      & Hetero                                                      & Batch                                                   & Comm   & \cite{el2018computational}              & Monolithic      & Hetero                                                   & Batch                                                   & Comp    \\ 
\hdashline
\cite{chen2019collaborative}   & Independent     & Hetero                                                   & Batch                                                   & Comm   & \cite{huang2018distributed}             & Independent     & Homo                                                      & Batch                                                   & ND      \\ 
\hdashline
\cite{deng2020optimal}         & Loosely-coupled & Hetero                                                   & Batch                                                   & ND     & \cite{min2019learning}                  & Monolithic      & Homo                                                      & ND                                                      & ND      \\ 
\hdashline
\cite{goudarzi2021distributed} & Modular         & Hetero                                                   & Batch                                                   & Hybrid & \cite{chen2018optimized}                & Independent     & Hetero                                                   & Batch                                                   & Comm    \\
\end{longtable}
\normalsize

\section{Environmental Architecture}
\label{Sec:environmentalArchitecture}
\begin{figure}[!t]
\centering 
\includegraphics[width=0.7\textwidth, height=6cm]{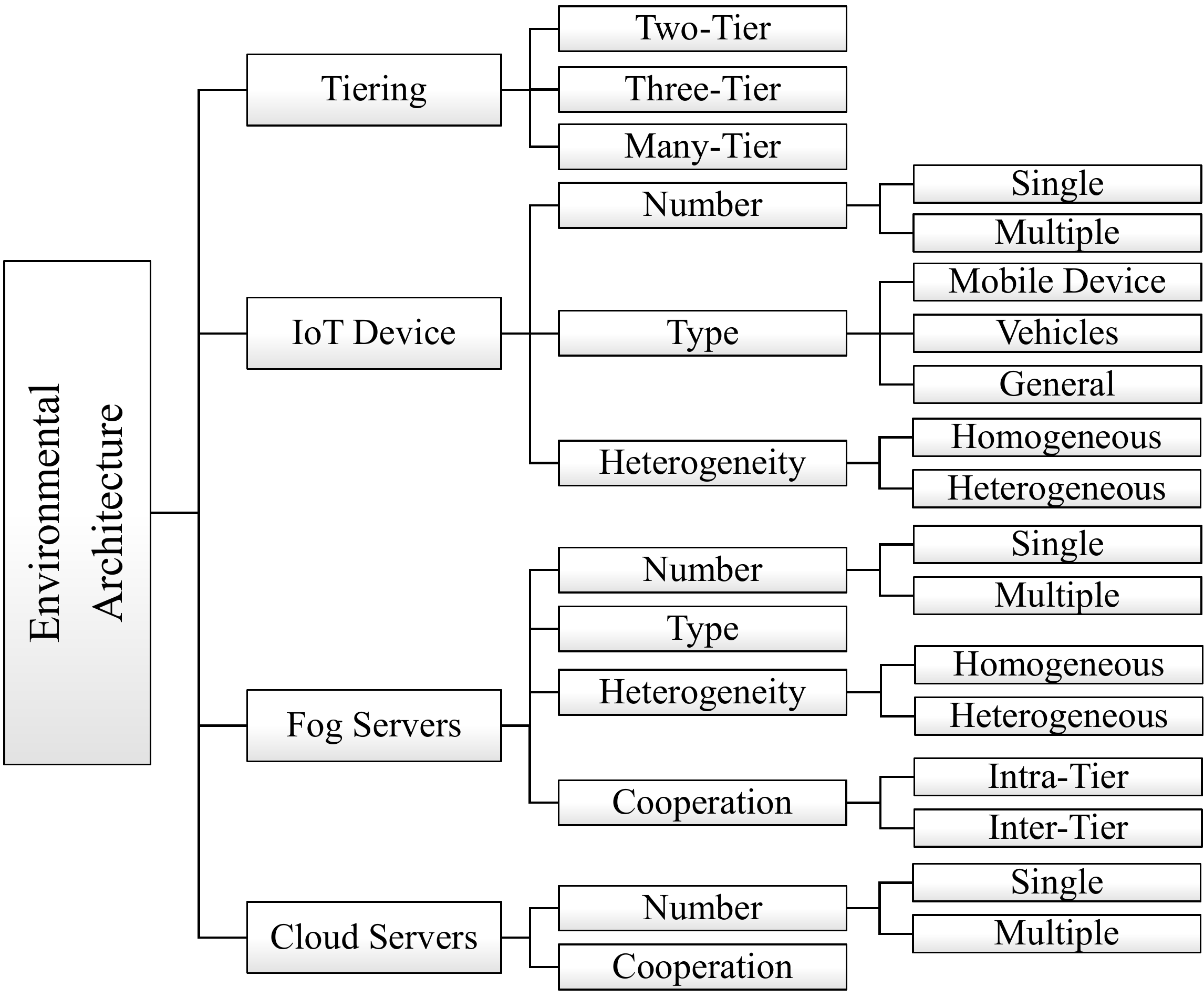}
\caption{Environmental architecture taxonomy}
\label{Fig:environmentalArchitectureTaxonomy}
\end{figure}

The configuration and properties of IoT devices and resource providers directly affect the complexity and dynamics of scheduling IoT applications. To illustrate, as the number of resource providers increases, heterogeneity in the system also grows as a positive factor, while the complexity of making a decision also increases that may negatively affect the process of making decisions. In this section, we classify the environmental architecture properties, as depicted in Fig.~\ref{Fig:environmentalArchitectureTaxonomy}, into the following categories:
\subsection{Tiering Model}
IoT devices and resource providers can be conceptually organized in different tiers based on their proximity to users and resources, described below:
\subsubsection{Two-Tier}
In this resource organization, IoT devices are situated at the bottom-most layer and resource providers are placed at the edge of the network in the proximity of IoT devices (i.e., Edge computing). Several works use two-tier resource organization such as \cite{maleki2021mobility,tang2020deep,huang2018distributed,liu2021task}. 
\subsubsection{Three-Tier}
Compared to two-tier model, this model also uses CSs at the highest-most layer to support edge resources (i.e., Fog computing). Several works considered three-tier model in the literature such as \cite{sarkar2021collaborative,pusztai2021pogonip,kishor2021task,ijaz2021energy}.      
\subsubsection{Many-Tier}
In many-tier resource organizations, IoT devices and CSs are situated at the bottom-most and highest-most tiers respectively, while FSs are placed in between through several tiers (i.e., hierarchical Fog computing). In the literature, several works have considered many-tier model such as \cite{goudarzi2021distributedmigration,gazori2020saving,rahbari2020task,maiti2019effective}. 
\subsection{IoT Devices}
IoT devices can play two roles; as service requester and/or resource provider. When they act as service requesters, still they can execute a portion of their tasks or components based on their available resources. Moreover, IoT devices can simultaneously play these different roles. We study the properties of IoT devices from the following perspectives:
\subsubsection{Number} The higher number of IoT devices (either as service requester or service provider), the higher complexity of the scheduling problem. Some works only consider single IoT device in the environment such as \cite{liu2021efficient,neto2018uloof,wang2019dynamic} while other works consider multiple IoT devices simultaneously such as \cite{ouyang2018follow,wang2019mobility,hoseiny2021pga}.
\subsubsection{Type} The type of IoT devices help us understand the amount of resources, capabilities, and constraints of these devices. The IoT devices used in the current literature can be broadly classified into three categories, namely 1) \textbf{mobile devices (MD)} which are mostly considered as smartphones or tablets \cite{deng2020optimal,shekhar2019urmila,pan2021multi}, 2) \textbf{Vehicles} \cite{wang2019delay,wang2020imitation,yang2019efficient}, and 3) \textbf{General} devices containing a set of IoT devices, ranging from small sensors to drones \cite{maleki2021mobility,goudarzi2021distributed,sheng2019computation}.
\subsubsection{Heterogeneity} We also study the resources of IoT devices and their request types, and classify proposals into 1) \textbf{heterogeneous} where IoT devices have various resources and different request types and sizes such as \cite{gazori2020saving,xu2019computation,lu2020optimization} or 2) \textbf{homogeneous} where the resources of IoT devices are the same or they have the same request type and size such as \cite{liu2021efficient,wang2019dynamic,yang2019joint}.
\subsection{Fog Servers (FSs)}
FSs usually act as resource providers for IoT devices. The environmental properties of FSs can be classified based on the following criteria:
\subsubsection{Number}
Similar to IoT devices, we classify the number of FSs in the environment into 1) \textbf{Single} and 2) \textbf{Multiple}. The complexity and dynamics of system in surveyed works that have considered only single FS such as \cite{huang2019deep,hu2021efficient,qiu2020distributed} is simpler to the works that have considered multiple FSs such as \cite{tao2021adaptive,yan2021two,farhadi2021service}. 
\subsubsection{Type}
The type of FSs acting as service provider in the Fog computing environment ranges from IoT devices with additional resources to resource-rich data centers. Several works have considered a specific type of FS and discuss their properties in their works such as 1) \textbf{Base Station (BS)} and \textbf{Macro-cell Station (MS)} \cite{chen2019collaborative,tao2021adaptive,wang2019dynamic}, 2) \textbf{femtocells} \cite{chen2018multi,goudarzi2019fog,gedawy2020ramos}, 3) \textbf{Rpi} \cite{fu2021adaptive} and 4) \textbf{access points (AP)} \cite{meng2019online,el2018computational}. Moreover, several works consider 5) \textbf{general} FSs containing a set of FSs with different types such as \cite{nan2018dynamic,gazori2020saving}.
\subsubsection{Heterogeneity}
We study the FSs' resources and classified works based on their heterogeneity into 1) \textbf{heterogeneous} and 2) \textbf{homogeneous} accordingly. Many works have considered heterogeneous resources for FSs \cite{asheralieva2019learning,goudarzi2021distributed,pan2021multi,hazra2021ceco} while some works consider homogeneous resources for FSs \cite{wang2021dependent,huang2019deep,lu2020edge}.  
\subsubsection{Cooperation} 
Compared to CSs, each FS has fewer resources and it may not be able to satisfy the requirements of IoT applications. Cooperation among FSs helps augmenting their resources and providing service for demanding IoT applications. We classify proposals based on their cooperation among FSs into 1) \textbf{intra-tier} where FSs of same tier collaborate to satisfy users' requests \cite{deng2020optimal,deng2021fogbus2,qi2019knowledge,deng2020optimal} and 2) \textbf{inter-tier} where FSs of different layers also collaborate for the execution of IoT one application \cite{goudarzi2021distributed,goudarzi2021distributedmigration}.
\subsection{Cloud Servers (CSs)}
The environmental properties of CSs can be classified based on the following criteria:
\subsubsection{Number}
The current literature based on the CSs' number can be divided into 1) \textbf{single} and 2) \textbf{multiple} categories. Majority of works only consider one CDC as resource provider (either as a central entity with aggregated resources or different number of VMs) to support FSs such as \cite{bashir2019resource,hoseiny2021pga,ijaz2021energy,kishor2021task}. In real-world environment, different CDCs are available which can provide services with different QoS for multiple applications. Some works such as \cite{goudarzi2021distributed,wang2020joint,goudarzi2020application,goudarzi2021resource} have considered multiple CDCs with heterogeneous CSs in the literature.
\subsubsection{Cooperation}
Among the works considered multi CDCs, we study either CSs from different CDCs are configured to collaboratively execute an IoT application or not. In the literature, some works such as \cite{goudarzi2021resource,goudarzi2020application,peng2021constrained} have considered collaborative multi CDCs scenarios.  
\subsection{Discussion}
In this section, we discuss the effects of identified environmental architecture's elements on the decision engine and describe the lessons that we have learned. Besides, we identify several research gaps accordingly. Table~\ref{Tab:EnvironmentalArchitecture} provides a summary of properties related to environmental architecture in Fog computing.
\subsubsection{\textbf{Effects on the decision engine}} The elements of environmental architecture affect the decision engine in various aspects, as briefly described below. 
\paragraph{1} 
Tiering: It represents the organization of end-users' devices and resources in the computing environment. Considering the properties of resources in different tiers, it helps find the most suitable deployment layer for the decision engine to efficiently serve IoT applications' requests with a wide variety of requirements. For example, to serve real-time IoT applications with low startup time requirements, the most suitable deployment layer in the three-tier model is the lowest-level Fog layer.
\paragraph{2} 
IoT devices: The number of IoT devices directly relates to the number of incoming requests to decision engines. It affects the admission control of decision engines. The type of IoT devices provides contextual information about the number of resources and intrinsic properties of the IoT devices that are important for the decision engine. For example, the MD type not only states that the IoT device does not have significant computing resources, but also presents that the device has mobility features. Thus, the IoT device type affects the advanced features of the decision engine, such as mobility, and also specifies whether the IoT devices can serve one or several tasks/modules of IoT applications or not. 
\paragraph{3} 
Fog and Cloud servers: The number of available servers directly affects the TC of the scheduling problem. Considering an application with $n$ number of tasks and $m$ possible candidate configuration per task, the TC of finding the optimal solution is $O(m^n)$. Hence, it directly affects the choice of placement technique and scalability feature of the decision engine. As the problem space increases, a suitable decision engine should be selected to solve the scheduling problem. Moreover, the type and heterogeneity of resources provide further contextual information for the decision-making, such as the number of resources, networking characteristics, and resource constraints, just to mention a few.
\subsubsection{\textbf{Lessons learned}} Our findings regarding the environmental architecture in the surveyed works are briefly described in what follows:
\paragraph{1} 
Almost 60\% of works consider the three-tier model and many-tier models for the organization of end-users and resources. Not only do these works consider real-time applications, but also some of them assume both real-time and computation-intensive applications, such as \cite{cai2021failure,stavrinides2019hybrid,goudarzi2021distributed}. This is mainly because these works use CSs as a backup plan for more computation-intensive applications or when the number of incoming IoT requests increases and the FSs cannot solely manage the incoming requests. Moreover, nearly 40\% of surveyed works assume a two-tier model for the organization of end-users and resources. These works mostly assume real-time workload type and communication-intensive applications for the deployment on Edge servers, such as \cite{bahreini2021vecman,cheng2021multi,hu2021efficient,qiu2020distributed}.  
\paragraph{2} 
In the surveyed works, almost 90\% of the works considered an environment with multiple IoT devices, while 10\% of works only focused on a single IoT device. When the number of IoT devices increases, the diversity of IoT applications and heterogeneity of their tasks also increase accordingly. Moreover, the greater number of works assume IoT devices as general devices with sensors, actuators, and diverse application requests. In contrast, some works targeted a specific IoT devices such as mobile devices and vehicles with almost 30\% and 10\% of proposals, respectively. These proposals studied other contextual information of targeted IoT devices in detail such as mobility \cite{wang2019dynamic}, energy consumption \cite{wang2020imitation}, and networking characteristics \cite{wang2019delay,qi2019knowledge}. Finally, about 90\% of works have studied IoT devices with heterogeneous properties and diverse application request types, which are the closest scenario to real-world computing environments.
\paragraph{3} 
Regarding Fog resources, almost 90\% of the proposals consider multiple FSs in the environment. However, only 40\% of the current literature has considered any cooperation model among FSs. There is a high probability that a single FS cannot solely manage the execution of several incoming resources due to its limited resources. Also, sending partial/complete applications' tasks to the Cloud may negatively affect IoT devices' response time and energy consumption, especially for real-time IoT applications. Thus, cooperation among FSs is of paramount importance that can lead to the execution of IoT applications with better performance and QoS. Considering the type of the FSs, about 60\% of the studied literature considered general FSs. The rest of the works studied a specific type of FSs and tried to involve their contextual information in the scheduling process of IoT applications, such as networking characteristics \cite{asheralieva2019learning}. Moreover, some works considered IoT devices can simultaneously play different roles in the computing environments (i.e., service requester and service provider) such as \cite{wang2020joint,wang2020imitation,zhu2018fog}. 
\paragraph{4} In the current literature, around 60\% of the works consider CSs as computing resources in the environment. However, only in 8\% of these works multiple Cloud service providers (i.e., multi-Cloud), their communication, and interactions are studied, such as \cite{goudarzi2020application,goudarzi2021distributed,peng2021constrained,wang2020joint}. 
\subsubsection{\textbf{Research Gaps}}
We have identified several open issues for further investigation, as discussed below:
\paragraph{1}
Hierarchical Fog computing (i.e., multi-tier) has not been thoroughly considered by researchers. Only a few works (almost 5\%) consider the organization of resources in the multi-tier environment, and most have focused on the heterogeneity of resources among different tiers. However, these works have not considered the heterogeneity of communication protocols and latency in multi-tier environments.    
\paragraph{2}
In the literature, several works have considered abstract CDC as a central unit with huge computing capacity \cite{mahmud2019quality}, while in reality, CDCs contain several CSs hosting computing instances. Such assumptions affect the computing and communication time in simulation studies.     
\paragraph{3}
One of the main advantages of Fog computing is providing heterogeneous FSs in IoT devices' vicinity to collaboratively serve applications. However, many works have not considered cooperation among FSs. In this case, due to the limited computing and communication resources of each FS and a large number of IoT requests, the serving FS may become a bottleneck which negatively affects the response time and QoS \cite{goudarzi2019fog}. Besides, in uncooperative scenarios, the overloaded FS forwards requests to CSs, incurring higher latency. Hence, cooperative Fog computing, associated protocols, and constraints require further investigation for different IoT application scenarios.
\scriptsize
\begin{longtable}{ccccccccccc}
\caption{Summary of existing works considering environmental architecture taxonomy \label{Tab:EnvironmentalArchitecture}}\\ 
\hline
\multirow{3}{*}{Ref}                             & \multicolumn{10}{c}{Environmental Architecture}                                                                                                                                                                                           \\* 
\cline{2-11}
                                                 & \multirow{2}{*}{Tiering} & \multicolumn{3}{c}{IoT Device} & \multicolumn{4}{c}{Fog Servers}                                                                                                       & \multicolumn{2}{c}{Cloud Servers}     \\* 
\cmidrule(lr){3-5}\cmidrule(lr){6-9}\cmidrule(r){10-11}
                                                 &                          & Number   & Type    & Hetero    & Number   & Type                                                      & Hetero & Coop                                                  & Number   & Coop                       \endfirsthead 
\cmidrule(lr){1-1}\cmidrule(lr){2-2}\cmidrule(r){3-5}\cmidrule(lr){6-9}\cmidrule(lr){10-11}
\cite{asheralieva2019learning}          & Two-Tier                 & Multiple & MD      & Hetero    & Multiple & BS, MS                                                    & Hetero & \scalebox{1.1}{\ensuremath \Circle}                                                & \scalebox{1.1}{\ensuremath \Circle}   & \scalebox{1.1}{\ensuremath \Circle}                     \\ 
\hdashline
\cite{bahreini2021vecman}               & Two-Tier                 & Multiple & Vehicle & Hetero    & Multiple & RSU                                                       & ND     & \scalebox{1.1}{\ensuremath \Circle}                                                & \scalebox{1.1}{\ensuremath \Circle}   & \scalebox{1.1}{\ensuremath \Circle}                     \\ 
\hdashline
\cite{chen2019collaborative}            & Three-Tier               & Multiple & MD      & ND        & Multiple & BS                                                        & Hetero & Intra                                                 & Single   & \scalebox{1.1}{\ensuremath \Circle}                     \\ 
\hdashline
\cite{deng2020optimal}                  & Three-Tier               & Multiple & MD      & Hetero    & Multiple & BS                                                        & Hetero & Intra                                                 & Single   & \scalebox{1.1}{\ensuremath \Circle}                     \\ 
\hdashline
\cite{goudarzi2021distributed}          & Many-Tier                & Multiple & General & Hetero    & Multiple & General                                                   & Hetero & \begin{tabular}[c]{@{}c@{}}Intra, Inter\end{tabular} & Multiple & \scalebox{1.1}{\ensuremath \CIRCLE}                   \\
\end{longtable}
\normalsize
\section{Optimization Characteristics}
\label{Sec:optimizationProblem}
\begin{figure}[!t]
\centering 
\includegraphics[width=0.5\textwidth, height=5cm]{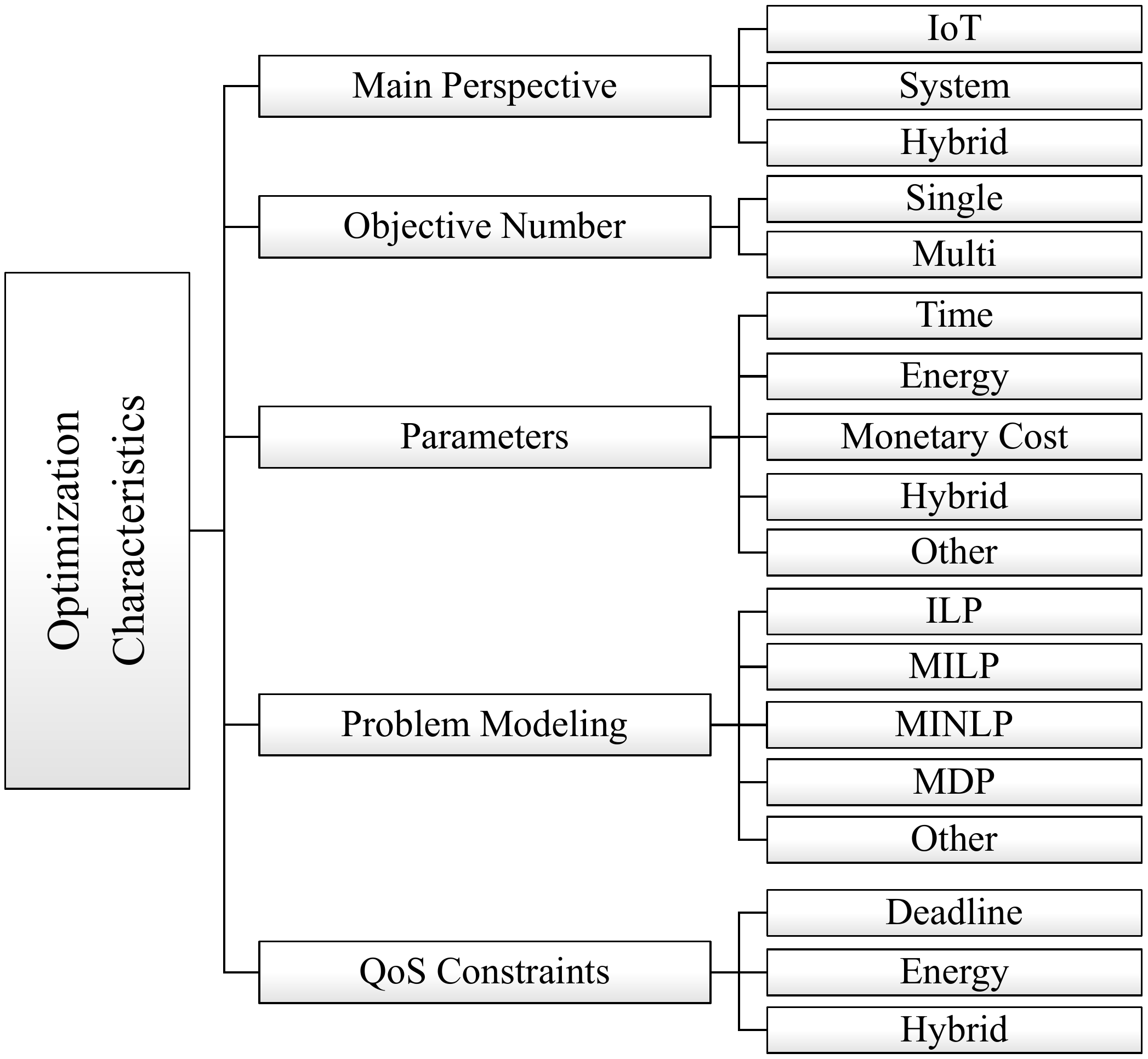}
\caption{Optimization characteristics taxonomy}
\label{Fig:optimizationCharacteristicsTaxonomy}
\end{figure}

Considering the application structure, environmental parameters, and the target objectives, each proposal formulates the problem of scheduling IoT applications in Fog computing. Optimization parameters directly affect the selection process and properties of suitable decision engines. Fig.~\ref{Fig:optimizationCharacteristicsTaxonomy} presents the principal elements in optimization characteristics, as described in what follows: 
\subsection{Main Perspective}
The proposals in the literature can be divided into three categories based on their main optimization goal, namely IoT devices/users, system, and hybrid, which are described in the following:
\subsubsection{IoT devices/users} The main perspective of several proposals is to satisfy the requirements of IoT applications such as minimizing their execution time and energy consumption of IoT devices, or improving user experience in terms of QoS and Quality of Experinece (QoE). Several works have considered IoT perspective for the optimization such as \cite{bahreini2021vecman,stavrinides2019hybrid,wang2021dependent,sami2020vehicular,sami2020vehicular,lu2020edge}. 
\subsubsection{System} The main perspective of this category is to improve the efficiency of resource providers such as minimizing their energy consumption, improving resource utilization, and maximizing the monetary profit \cite{farhadi2021service,chen2018multi,bahreini2021mechanisms,hoseiny2021pga}. Hence, these works often assume IoT devices with very limited limited computational resources that transfer sensed data to the surrogate servers for processing and storage.
\subsubsection{Hybrid} Some proposals targeted optimizing the parameters of both IoT devices/users and resource providers, referred to as hybrid optimization \cite{huang2018distributed,xiong2020resource,maia2021improved,nguyen2021multi}. In these works, IoT devices have some computational resources to serve their partial/complete tasks. However, they may send their requests to other surrogate servers if overall global parameters of IoT devices and systems can be optimized.
\subsection{Objective Number} According to the number of main optimization objectives of each proposal, we classify the current literature into 1) \textbf{single objective} and 2) \textbf{multi objective} proposals. Multi objective proposals consider several parameters to simultaneously optimize them, incurring higher complexity. In the literature, several proposals targeted single objective optimization such as \cite{liu2021task,sami2020vehicular,zhu2018fog,huang2019deepDigital}, while other proposals try to optimize several parameters such as \cite{xu2019computation,goudarzi2021distributedmigration,natesha2021adopting,hazra2021ceco}.
\subsection{Parameters}
According to the main objectives and the nature of optimization parameters in the literature, we categorize these parameters into the following categories:
\subsubsection{Time} 
One of the most important optimization parameters is the execution time of IoT applications. Minimizing the execution time of IoT applications provides users with a better QoS and QoE. This category contains any metrics related to time such as response time, execution time, and makespan used in the literature such as  \cite{han2019ondisc,wang2020fast,deng2021fogbus2,peixoto2021hierarchical}.
\subsubsection{Energy} 
IoT devices are usually considered as battery-limited devices. Hence, minimizing their energy consumption is one of the most important optimization parameters. Besides, energy consumption from FSs' perspective is two-fold. First, some FSs, similar to IoT devices, are battery-constrained, making optimizing the energy consumption of FSs an important challenge. Second, from the system perspective, the energy consumption of FSs should be minimized to reduce carbon emissions. This category contains any proposals considered energy consumption as an optimization parameter either from IoT devices or system perspectives such as \cite{qiu2020distributed,ijaz2021energy,huang2019deep,bahreini2021vecman}.
\subsubsection{Monetary Cost} 
This category studies the proposals that have considered the monetary aspects either from IoT users (i.e., minimizing monetary cost) or system perspectives (i.e., increasing monetary profit) \cite{bahreini2021mechanisms,nan2018dynamic,nan2018dynamic,pusztai2021pogonip,ma2021towards}.
\subsubsection{Other}
Some works have considered other optimization parameters such as the number of served requests, system utility, and resource utilization, just to mention a few, such as \cite{farhadi2021service,sami2021demand,chen2018multi,goudarzi2019fog}.
\subsubsection{Hybrid}
Several works also have considered a set of optimization parameters, referred as hybrid. These works use any combination of above-mentioned parameters simultaneously \cite{goudarzi2021distributed,goudarzi2020application,wang2021dependent,hazra2021ceco}.
\subsection{Problem Modeling}
Considering the main goal and optimization parameters, the optimization problem can be modeled/formulated. Considering surveyed literature in terms of the problem modeling approach, we classify the works into the following categories:
\subsubsection{Integer Linear Programming (ILP)}
It is a problem type where the variables and constraints are all integer values, and the objective function and equations are linear. Several works have used ILP for problem modeling such as \cite{neto2018uloof,wang2020service,mouradian2019application,ma2021towards}.
\subsubsection{Mixed Integer Linear Programming (MILP)}
In these problems, only some of the variables are constrained to be integers, while other variables are allowed to be non-integers. Also, the objective function and equations are linear. Several works have modelled their problem as an MILP such as \cite{peng2021constrained,xu2019computation,huang2018distributed,liu2021efficient}. 
\subsubsection{Mixed Integer Non-Linear Programming (MINLP)}
It refers to problems with integer and non-integer variables and non-linear functions in the objective function and/or the constraints. Several works such as \cite{maleki2021mobility,zhou2020online,yang2020computation,cheng2021multi} have used MINLP to present their optimization problems. 
\subsubsection{Markov Decision Process (MDP)}
It provides a mathematical framework to model and analyzes problems with stochastic and dynamic systems. Several works have used MDP to model scheduling problem in Fog computing such as \cite{goudarzi2021distributed,sami2021demand,wang2021dependent,asheralieva2019learning}.  
\subsubsection{Other}
There are also some other optimization modeling approaches in the literature of scheduling applications in Fog computing such as game theory \cite{hong2019multi}, lyapunov \cite{ouyang2018follow,chen2018multi}, and mixed integer programming \cite{sarkar2021collaborative,aburukba2021heuristic}.
\subsection{QoS Constraints} The formulated optimization problem usually contains several constraints, incurring higher complexity compared to unconstrained problems. In this work, we classify techniques based on the QoS-related constraints applied to the main formulated problem into 1) \textbf{deadline} such as \cite{wang2020imitation,gedawy2020ramos,tian2021user}, 2) \textbf{energy} such as \cite{yan2021two,badri2019energy}, and 3) \textbf{hybrid} (i.e., any combination of deadline, energy, and cost) such as \cite{ouyang2018follow,hazra2021ceco,cheng2021multi}.     
\subsection{Discussion}
In this section, we discuss the effects of identified optimization characteristics' elements on the decision engine and describe the lessons that we have learned. Besides, we identify several research gaps accordingly. Table~\ref{Tab:OptimizationCharacteristics} provides a summary of characteristics related to optimization problems in Fog computing.
\subsubsection{\textbf{Effects on the decision engine}} The optimization characteristics affect the decision engine in various aspects, as briefly described below. 
\paragraph{1} 
Objective number and parameters: Simultaneous optimization of multi-objective problems usually incur higher complexity for the decision engine. Also, when the number of key parameters in a multi-objective scheduling problem increases, finding the best parameters' coefficients becomes a critical yet challenging step.
\paragraph{2} 
Problem modeling: It can affect the choice of placement technique as some specific algorithms and techniques can be used to solve the scheduling problem. For example, several traditional LP and ILP tools and libraries exist to solve LP and ILP scheduling problems.
\paragraph{3} 
QoS constraints: They incur hard or soft constraints and limitations on the main objective/objectives of the scheduling problem, which intensify the complexity of the scheduling problem. The decision engine should satisfy these constraints either using classic Constraint Satisfaction Problem (CSP) techniques or using customized approaches. 
\subsubsection{\textbf{Lessons learned}} Our findings regarding the optimization characteristics in the surveyed works are briefly described in what follows:
\paragraph{1} 
The main perspective of optimization for almost 75\% of works is IoT, while for the rest of the works is hybrid and system by 15\% and 10\%, respectively. The main perspective element affects how some metrics are evaluated. For example, when evaluating the energy consumption in the IoT perspective, the energy consumed by the surrogate servers for the execution of tasks is overlooked. However, in the system and hybrid perspectives, the energy consumption of all resource providers and all entities in the systems are evaluated, respectively.
\paragraph{2}
Considering objective numbers in the optimization problem, the works are almost equally divided into single and multiple objective numbers. Overall, the majority of works studied time and/or energy as their main optimization metrics. While the works with an IoT perspective follow the same trend for the optimization metrics, the proposals with a system perspective almost consider the cost as their main optimization parameter. Also, the hybrid perspective proposals often consider a combination of time, energy, and/or cost as their main optimization metrics.
\paragraph{3}
In problem modeling, the greater number of works have used either MDP or MINLP (each with roughly 25\% of proposals) to formulate their problem. Also, some works initially had modeled their work as MINLP and then defined the MDP accordingly, such as \cite{cheng2021multi,huang2019deepDigital}. The rest of the works have used MILP (almost 15\%), ILP (almost 15 \%), and other optimization modeling approaches.
\paragraph{4} 
Almost 25\% of works defined single or multiple QoS constraints for their problem, among which 90\% have considered a single constraint, and the rest went for two QoS constraints. Among the QoS constraints, the deadline by 90\% is the most used constraint in all works. 
\subsubsection{\textbf{Research Gaps}}
We have identified several open issues for further investigation that are discussed below:
\paragraph{1} 
The main part of works in the literature either consider optimization problems from IoT devices/users. However, only a few works have considered IoT and system perspectives simultaneously (i.e., hybrid). Optimizing either of these perspectives can negatively affect other perspectives. To illustrate, when the principal target is minimizing the energy consumption of IoT devices, the majority of components or tasks are placed at FSs or CSs. However, it may negatively affect the energy consumption of resource providers and even increase the aggregated energy consumption in the environment. Hence, further investigation on hybrid optimization perspectives and mutual effects of different perspectives is required.
\paragraph{2}
The cooperation among the resource providers (i.e., FSs, CSs) is an essential factor in offering higher-quality services. Proposals in the system and hybrid perspectives can also consider other metrics such as trust and privacy index for resource providers and study how they affect the overall performance.
\paragraph{3} 
QoS constraints are set to guarantee a minimum service level for end-users. In current literature, most of proposals have focused on the deadline as the constraint. However, several other parameters such as privacy, security, and monetary cost and their combination as hybrid QoS constraints are not studied. 
\section{Decision Engine Characteristics}
\label{Sec:decisionEngines}
\begin{figure}[!t]
\centering 
\includegraphics[width=0.7\textwidth, height=7cm]{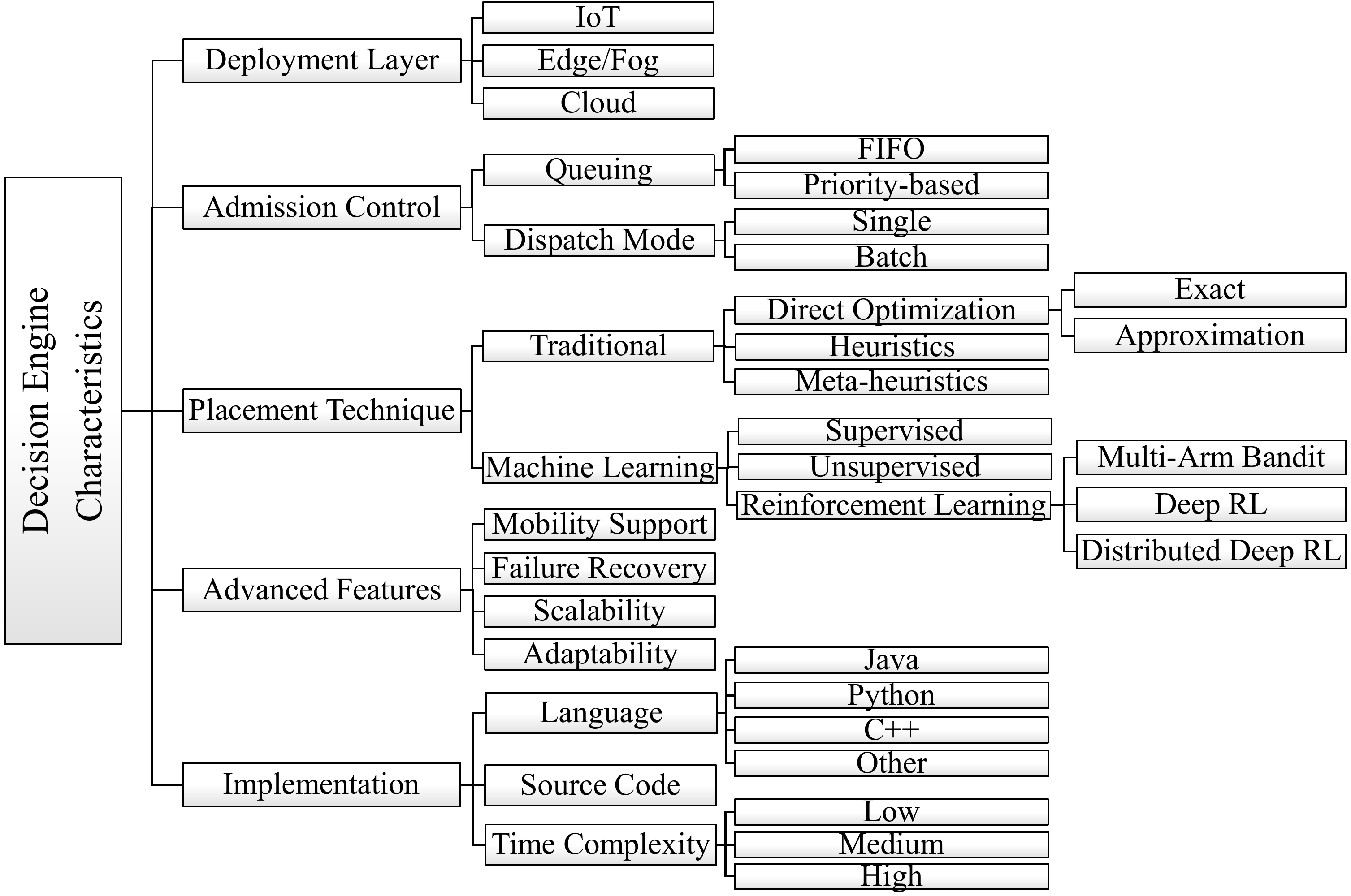}
\caption{Decision engine taxonomy}
\label{Fig:decisionEngineTaxonomy}
\end{figure}

The requirements of IoT applications in Fog computing can be satisfied if incoming IoT requests can be accurately scheduled based on the characteristics of application structure, environmental architecture, and optimization problems by the decision engine. The main responsibilities of the decision engine are organizing received IoT requests and solving the optimization problem through a placement decision while considering contextual information. Fig.~\ref{Fig:decisionEngineTaxonomy} presents
the main elements in decision engine, as described in what follows:
\subsection{Deployment Layer}
Decision engines can be deployed on servers at different layers unless the servers do not have sufficient resources to host them. Based on the deployment layer of the decision engine, the current literature can be classified into: 
\newpage
%\begin{landscape}
\scriptsize
%\refstepcounter{table}
\begin{longtable}{cccccc||cccccc} 
\caption{Summary of existing works considering optimization Characteristics taxonomy \label{Tab:OptimizationCharacteristics}}\\
\hline
\multirow{2}{*}{Ref}                    & \multicolumn{5}{c||}{Optimization Characteristics}                                                    & \multirow{2}{*}{Ref}                             & \multicolumn{5}{c}{Optimization Characteristics}                                                                                                                                                                                                                                                              \\* 
\cline{2-6}\cline{8-12}
& \begin{tabular}[c]{@{}c@{}}Main\\Persp\end{tabular} & \begin{tabular}[c]{@{}c@{}}Object\\Number\end{tabular} & Metrics                                                                 & \begin{tabular}[c]{@{}c@{}}Prob\\Model\end{tabular} & \begin{tabular}[c]{@{}c@{}}QoS\\Const\end{tabular}      &                                                  & \begin{tabular}[c]{@{}c@{}}Main\\Persp\end{tabular} & \begin{tabular}[c]{@{}c@{}}Object\\Number\end{tabular} & Metrics                                                                  & \begin{tabular}[c]{@{}c@{}}Prob\\Model\end{tabular} & \begin{tabular}[c]{@{}c@{}}QoS\\Const\end{tabular}         \endfirsthead 
\hline
\cite{asheralieva2019learning} & System                                              & Single                                                 & Cost                                                                    & MDP                                                    & \scalebox{1.1}{\ensuremath \Circle}                                                 & \cite{mahmud2019context}                & IoT                                                 & Single                                                 & Time                                                                     & ILP                                                    & Deadline                                                   \\ 
\hdashline
\cite{bahreini2021vecman}      & IoT                                                 & Single                                                 & Energy                                                                  & MINLP                                                  & \scalebox{1.1}{\ensuremath \Circle}                                                 & \cite{el2018computational}              & IoT                                                 & Multiple                                               & \begin{tabular}[c]{@{}c@{}}Time,\\Cost\end{tabular}                      & MINLP                                                  & Deadline                                                   \\
\end{longtable}
\normalsize
\subsubsection{IoT Layer}
The IoT devices usually are considered as resource-limited and battery-constrained devices. Hence, decision engines running on IoT devices should be very lightweight even with compromising the accuracy. In the literature, several works such as \cite{neto2018uloof,tang2020deep,meng2019online,huang2019deepDigital} deployed decision engines at the IoT layer.  
\subsubsection{Fog/Edge Layer}
Distributed FSs with sufficient resources situated in the proximity of IoT devices are the main deployment targets for the decision engines. They provide low-latency and high-bandwidth access to decision engines for IoT devices. Majority of works such as \cite{pan2021multi,goudarzi2021distributed,min2019learning,kimovski2021mobility} deployed the decision engines in Edge/Fog Layer.
\subsubsection{Cloud Layer}
CSs are potential targets for the deployment of decision engines. Although the access latency to CSs is higher, they provide high availability, making them a suitable deployment target where FSs are not available or when IoT applications are insensitive to higher startup time. Some works such as \cite{el2018computational,shekhar2019urmila} considered cloud layer for the deployment of decision engines. 
\subsection{Admission Control}
The admission control presents the behavior of decision engines when new requests arrive. It denotes how the new requests are queued and organized by the dispatching module for placement.
\subsubsection{Queuing} 
Decision engine may use different queuing policies when incoming IoT requests arrives. Based on queuing policy, we classify works into 1) \textbf{First-in-First-Out (FIFO)} such as \cite{gedawy2020ramos,tang2020deep,goudarzi2021distributed} and 2) \textbf{Priority-based} where incoming requests are sorted based on their priority (e.g., deadline) \cite{han2019ondisc,tian2021user,hoseiny2021pga}.
\subsubsection{Dispatching Mode}
The dispatching module forwards requests from input queue to the placement module. Based on the selection policy of dispatching module, current literature can be classified to 1) \textbf{single} model where only one task at a time is dispatched for placement \cite{ho2020joint,xie2019mobility,cheng2021multi} and 2) \textbf{batch} model where a set of tasks are forwarded to placement module \cite{goudarzi2020application,abd2021advanced,nguyen2021multi}. 
\subsection{Placement Technique}
Placement technique is the actual algorithm used to solve the optimization problem. Each placement algorithm has its advantages and disadvantages. Hence, it should be carefully selected based on the properties and dynamics of applications, users, environment, and deployment layer. We classify placement techniques based on their approach to find the solution into two broad categories: 
\subsubsection{Traditional} In this approach, the programmer/designer defines the required logic of policies for the placement technique. The traditional placement technique can be further divided into three subcategories:
\paragraph{1} \textbf{Direct Optimization}: In this category, the optimization problem will be solved using classical optimization tools either using 1) \textbf{Exact} approach to find the optimal solution such as \cite{chen2019collaborative,deng2020optimal} or 2) \textbf{Approximation} approach to find a near optimal solution such as \cite{maleki2021mobility,wang2020service,badri2019energy}. 
\paragraph{2} \textbf{Heuristics}: These algorithms are a set of typically problem-dependent algorithmic steps to find a feasible solution for the problem. Heuristics usually scale well as their TC is low while they do not guarantee finding the optimal solution of the problem \cite{bahreini2021vecman,liu2021efficient,farhadi2021service,bashir2019resource}.
\paragraph{3} \textbf{Meta-heuristics}: 
Meta-heuristics are composed of several advanced heuristics and typically are problem-independent, such as Genetic Algorithm (GA) and Simulated Annealing (SA). Although these algorithms usually perform better than heuristics, similarly they cannot guarantee to find the optimal solution. Several works such as  \cite{goudarzi2020application,wang2020joint,ghanavati2020energy,mouradian2019application} used meta-heuristics.
\subsubsection{Machine Learning (ML)}
ML is a family of algorithms that can learn the required policies for placement techniques from historic data. ML algorithms scale reasonably well, however, they require accurate and ideally large samples of historic data. The ML-based placement techniques can be further divided into three subcategories:
\paragraph{1} \textbf{Supervised Learning}: These algorithms learn by using labeled data as its input. Type of problems are regression and classification, and some of the algorithms are linear regression and logistic regression. Some works in the current literature used supervised ML for the placement technique such as \cite{yang2020computation,yan2021two,huang2019multi}.
\paragraph{2} \textbf{Unsupervised Learning}: These algorithms are trained using unlabelled data, contrary to supervised ML, without any guidance. Some unsupervised algorithms are K–Means and fuzzy C–Means. Some works in the current literature used unsupervised ML for the placement technique such as \cite{sheng2019computation,xie2019mobility,selimi2019lightweight}.
\paragraph{3} \textbf{Reinforcement Learning (RL)}: In these algorithms, agent/agents learn the required policy for placement technique by interaction with an uncertain and potentially complex environment. It does not require pre-defined data, and type of problems are exploitation or exploration. The current RL-based literature in scheduling IoT applications can be divided into 1) \textbf{Multi-Armed Bandit (MAB)} which are among the simplest RL problems such as \cite{zhou2020online,tao2021adaptive}, 2) \textbf{Deep RL (DRL)} where deep learning is used in RL such as \cite{asheralieva2019learning,tang2020deep,min2019learning}, and 3) \textbf{Distributed DRL} where several agents works collaboratively in a distributed manner for efficient learning such as \cite{goudarzi2021distributed,tuli2020dynamic,cheng2021multi}.
\subsection{Advanced Features}
To fully utilize the potential of the Fog computing paradigm, several advanced features can be augmented with decision engines to capture high dynamics of this paradigm, described below:
\subsubsection{Mobility Support} A significant number of IoT devices are moving entities (e.g., vehicles), requiring connected service through their path. So, decision engines should manage the migration process of application components and find suitable surrogate servers accordingly. Several works such as \cite{goudarzi2021distributedmigration,wang2019dynamic,badri2019energy,mouradian2019application} address mobility and migration management challenges alongside scheduling IoT applications.    
\subsubsection{Failure Recovery}
In highly dynamic systems such as Fog computing, failure may happen due to software or hardware-related issues. So, application components faced with failure should be re-executed. Some works consider failure recovery mechanisms in their decision engines such as \cite{goudarzi2020application,cai2021failure,goudarzi2021distributedmigration}. 
\subsubsection{High Scalability}
As a large number of IoT devices and servers exist in the Fog environment, mechanisms and algorithms used in the decision engine should be highly scalable and provide well-suited performance when the system size grows. Several works have studied the scalability feature of their techniques when the number of IoT applications and servers increases or discussed how their distributed techniques work efficiently in large systems such as \cite{hu2021efficient,fu2021adaptive,lu2020edge}. 
\subsubsection{High Adaptability}
This feature ensures that the decision engine dynamically captures the contextual information (i.e., application, environment, etc), and updates the policies of placement techniques accordingly. In Fog literature, several works such as \cite{gazori2020saving,goudarzi2021distributed,zhang2019task,pusztai2021pogonip} offer solutions with high adaptability.
\subsection{Implementation}
The implementation characteristics of decision engines are studied based on the following criteria:
\subsubsection{Language}
Different programming languages are used for the implementation of decision engines, while the majority have used Python \cite{ma2021towards,goudarzi2021distributed}, Java \cite{shekhar2019urmila,goudarzi2020application}, and C++ \cite{bahreini2021mechanisms,stavrinides2019hybrid}.
\subsubsection{Source Code}
Open-source decision engines help researchers to understand the detailed implementation specifications of each work, and minimize the reproducibility effort of decision engines, especially for comparison purposes. Some works such as \cite{deng2021fogbus2,wang2021dependent,qiu2020distributed} have provided the source code repository of their decision engines. 
\subsubsection{Time Complexity (TC)}
TC of each placement technique presents the required time to solve the optimization problem in the worst-case scenario. It directly affect the service startup time and the decision overhead of each technique. Based on the current literature, we classify the TC into 1) \textbf{Low} the solution of optimization problem can be obtained in polynomial time where the maximum power of variable is equal or less than two (i.e., $O(n^2)$) \cite{asheralieva2019learning,cai2021failure,goudarzi2021distributed}, 2) \textbf{medium} where the time complexity is polynomial and the maximum power is less than or equal to 3 (i.e., $O(n^3)$) \cite{bahreini2021vecman,liu2021efficient,peng2021constrained}, and 3) \textbf{High} for exponential TC and polynomials with high maximum power \cite{chen2019collaborative,deng2020optimal,tian2021user}.    
\subsection{Discussion}
In this section, we describe the lessons that we have learned regarding identified elements in decision engine characteristics of the current literature. Besides, we identify several research gaps accordingly. Table~\ref{Tab:DecisionEngine} provides a summary of decision engines-related characteristics in Fog computing.
\subsubsection{\textbf{Lessons learned}} Our findings regarding the decision engine characteristics in the surveyed works are briefly described in what follows:
\paragraph{1} 
Almost 85\% of surveyed works deployed the decision engine at the Edge layer in the proximity of IoT devices. Since the Edge servers can be accessed with lower latency and higher access bandwidth, deployment of decision engines at the Edge can reduce the startup of IoT applications. However, the Edge devices should have sufficient resources to run the decision engine. Some proposals (about 10\%) also deployed the decision engine on IoT devices. Deployment of a decision engine on IoT devices provides more control for IoT devices, especially mobile ones. It eliminates the extra overhead of communication with surrogate servers for making a decision. However, IoT devices often have very limited resources that are incapable of running powerful decision engines.
\paragraph{2}
The queuing is an important element in the admission control that almost 80\% of the works have not studied. Since most of works have considered several IoT devices in the environment, several IoT requests may arrive in each decision time-slot with a high probability. Hence, different queuing models can dramatically affect the decision engine performance and the QoS of end-users. FIFO and priority queue share the same proportion of proposals among the works that mentioned their queuing policy. Also, in priority-based queuing, almost all works have considered the deadline of applications or tasks as their main priority metric. Moreover, for the policy of dispatching module, about 75\% of works selected single dispatching while 25\% of works studied batch dispatching policy. Since different IoT requests may arrive in the same decision time-slot, the batch dispatching policy helps study the mutual effects of IoT applications with diverse resource requirements in the placement decision. 
\paragraph{3} 
The traditional placement techniques are used in almost 60\% of the proposals, while the ML-based placement techniques are studied in the rest of the works (almost 40\%). However, the number of ML-based placement techniques has significantly increased in recent years. In traditional placement techniques, direct optimization, heuristics, and meta-heuristics share the same proportion of proposals. Also, in meta-heuristics techniques, the majority of works used population-based meta-heuristics, especially different variations of the GA. In the ML-based techniques, the majority of proposals have used RL-based techniques (almost 70\%), specially DRL. Moreover, in the DRL techniques, the larger number of works used centralized DRL techniques such as DQN. However, the exploration and convergence rate of centralized DRL techniques are very slow. Thus, several studies have recently been conducted to adapt distributed DRL (i.e., DDRL) techniques for resource management in Edge/Fog computing environments, such as \cite{goudarzi2021distributed,hu2021efficient,lu2020edge}, to improve the exploration cost and convergence rate of the DRL techniques. 
\paragraph{4} 
In advanced features, almost 25\% of proposals embedded different mechanisms (i.e., traditional or ML-based) for the mobility management of IoT devices and migration of applications' constituent parts. Also, about 25\% of studied works, mostly ML-based techniques, offer high adaptability features in their decision engine. However, traditional works often neglect to provide different mechanisms to support high adaptability. This is mainly because the scheduling policies are not statically defined in ML-based techniques. Hence, as the environmental or application properties change, the policies can be learned and updated accordingly. However, in the traditional scheduling techniques, updating the scheduling policies according to dynamic changes in environmental or application properties is very costly and time-consuming. Almost 20\% of proposals studied different mechanisms to support high scalability feature, either using ML-based techniques or traditional approaches. In advanced features, the failure recovery mechanisms and techniques in scheduling are not well-studied and only a few works embedded these mechanisms in their scheduling techniques.
\paragraph{5}
Considering the implementation of the techniques, almost 50\% of the works mentioned their employed programming language. Java and python programming language are the most-employed programming language and are almost equally used in different proposals. However, Python is mainly used for ML-based techniques and direct optimization techniques, while Java is mostly used to implement traditional decision engines. Moreover, only about 10\% of proposals shared their open-source repositories with researchers and developers among the surveyed works. Finally, about 65\% of proposals discussed the TC of their works, among which almost 80\% proposed decision engines with low TC while some proposals (almost 10\%) went for medium TC and few works (almost 10\%) proposed decision engines with high TC. The high TC proposals are among the direct optimization category of traditional approaches. While these high TC proposals cannot be currently adapted to large-scale Edge and Fog computing environments, they can find the optimal solution in small-scale problems. Hence, they can be used as a reference for the evaluation of other proposals.  
\subsubsection{Research Gaps}
We have identified several open issues for further investigation that are discussed below:
\paragraph{1} 
The admission control concept in terms of different queuing, dispatching, and their mutual effect is not well studied in the current literature. Also, the greater number of works consider a single task dispatching model and overlook batch placement of applications, especially for applications with dependent tasks.  
\paragraph{2} 
While traditional placement techniques (e.g., heuristics, meta-heuristics) are studied well in the literature, ML-based techniques are still in their infancy. Due to the lack of a large number of datasets, supervised and unsupervised ML have not been thoroughly considered. Also, the majority of employed RL techniques are centralized approaches, neglecting collaborative learning of multiple distributed agents for better efficiency and lower exploration costs.
\paragraph{3} 
Although all servers and devices are prone to failures, among advanced features, failure recovery mechanisms, algorithms, and their integration with the placement technique is the least-studied concept. Even the best placement techniques cannot complete their process in real-world scenarios unless a suitable failure recovery mechanism is embedded.
\paragraph{4} 
In the surveyed works, there is no proposal to study all the four identified elements in the advanced features (i.e., mobility, failure recovery, scalability, and adaptability) and describe the behavior and mutual effects of these elements on each other and decision engine.
\paragraph{5}
Among the studied literature, none of the works has studied the privacy problem from different perspectives, such as end-users' data privacy, resource providers' privacy, and the decision engine's mechanisms for improving privacy.

\section{Performance Evaluation}
\label{Sec:performanceEvaluation}
Different approaches and metrics have been used by the research community to evaluate the performance of their techniques. Identifying and studying these parameters helps to select the best approach and metrics for the implementation of new proposals and fair comparisons with other techniques in the literature. Fig.~\ref{Fig:performanceEvaluationTaxonomy} presents a taxonomy
and the main elements of performance evaluation, described below:
\subsection{Approaches}
\begin{figure}[!t] 
	\centering 
	\begin{minipage}[t]{0.45\textwidth}
		\centering
		\includegraphics[height=4cm]{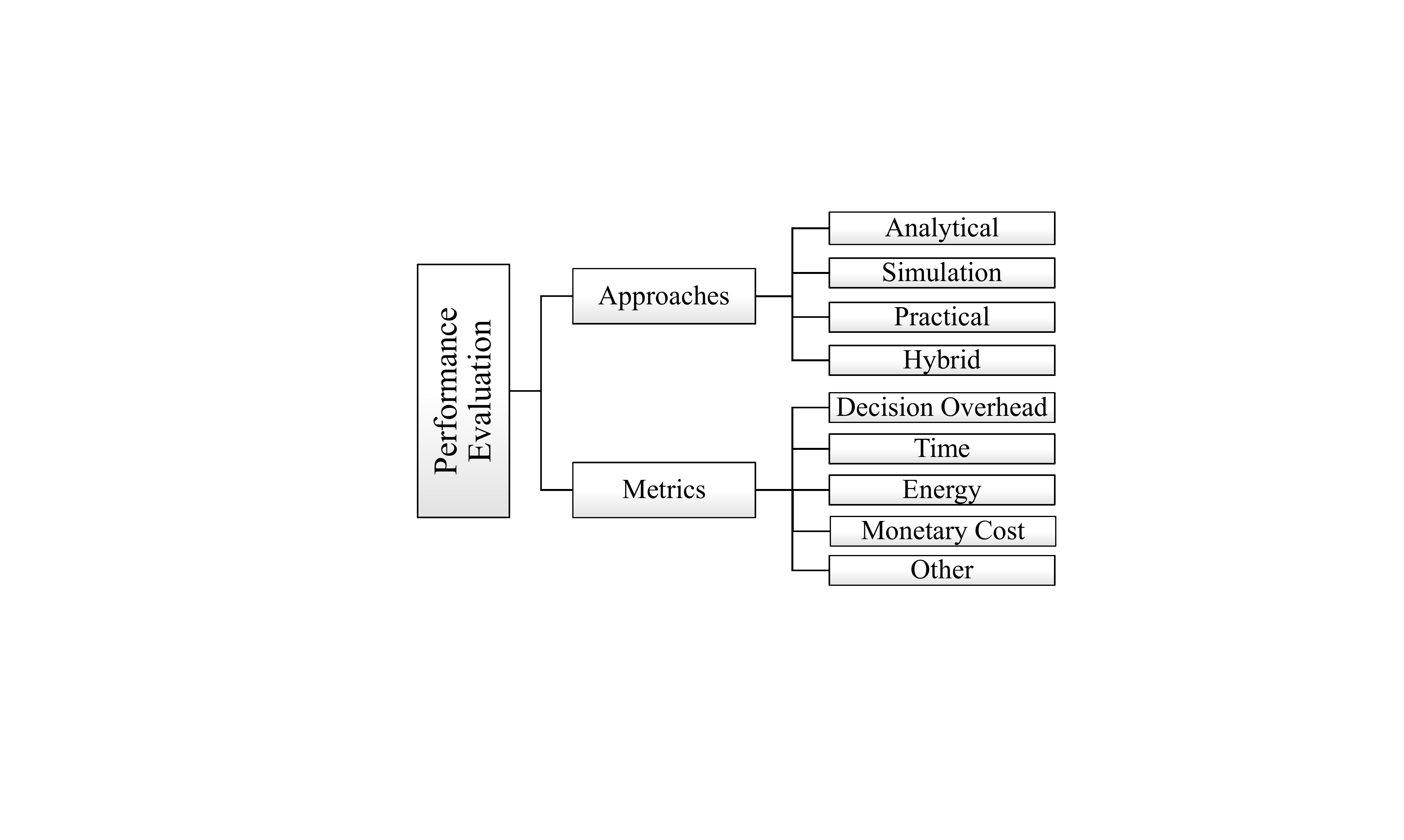} 
		\caption{Performance evaluation taxonomy}
		\label{Fig:performanceEvaluationTaxonomy}
	\end{minipage} 
	\hspace{1cm} 
	\begin{minipage}[t]{0.45\textwidth}
		\centering 
		\includegraphics[height=4cm]{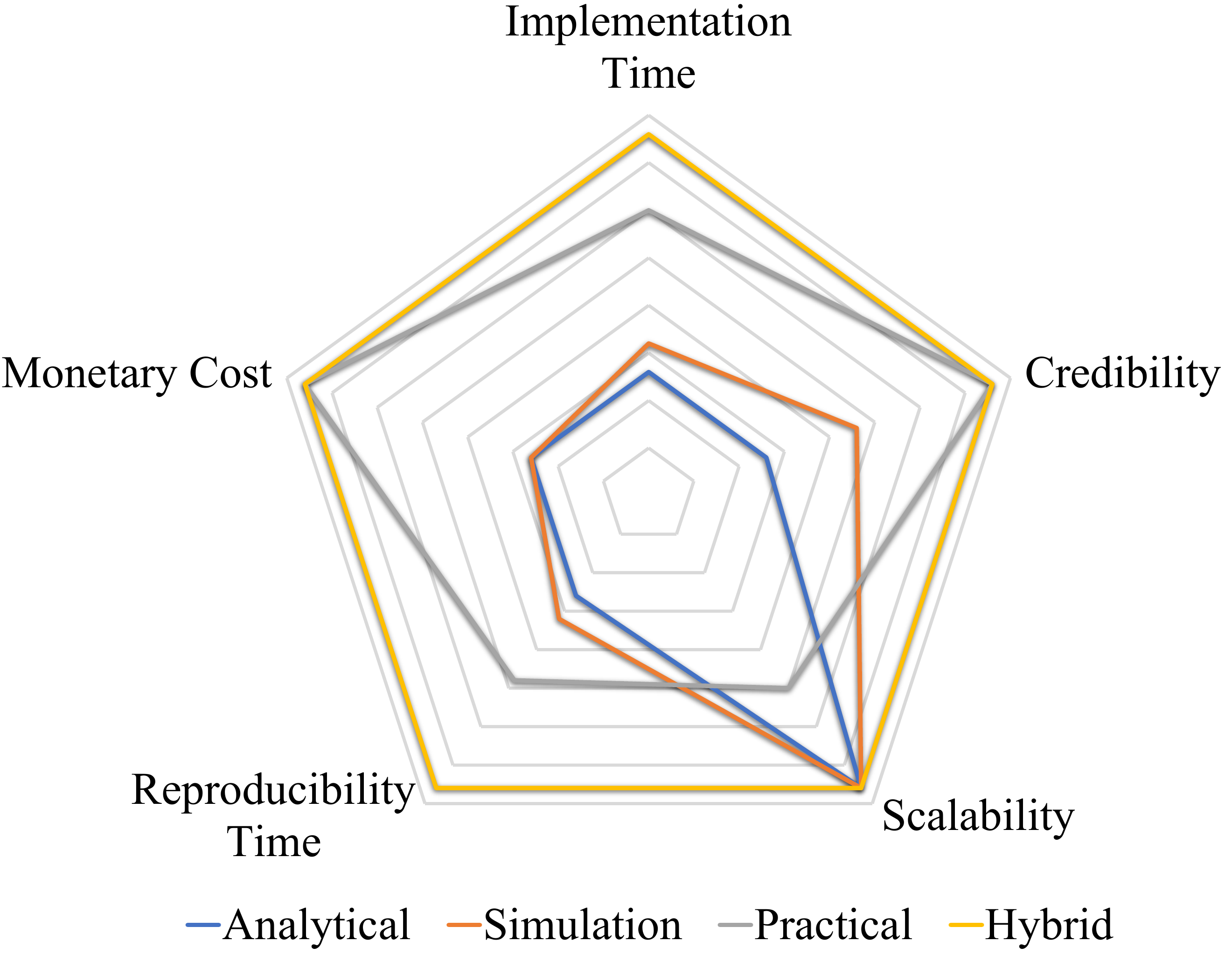} 
		\caption{Performance evaluation approaches}
		\label{Fig:performanceApproaches}
	\end{minipage} 
\end{figure}

The performance evaluation approaches can be divided into four categories, namely analytical, simulation, practical, and hybrid. There are different important aspects to consider when selecting an approach for the evaluation of proposals, such as credibility, implementation time, monetary cost, reproducibility time, and scalability. Fig.~\ref{Fig:performanceApproaches} presents a qualitative comparison of different approaches used in performance evaluation.
\subsubsection{Analytical}
One of the popular approaches for the evaluation of different proposals is analytical tools. Usually, the implementation time, reproducibility time, and monetary cost of analytical tools are low, and scalable experiments can be executed. However, the credibility of such experiments is low since the dynamics of resources, applications, and environment cannot be fully captured and tested. Matlab is among the most popular tools that is either used directly \cite{yang2020computation,yang2020computation,farhadi2021service} or integrated with some other libraries such as Sedumi\footnote{https://github.com/sqlp/sedumi} \cite{sarkar2021collaborative}. Also, C++ based analytical tools have been used in the literature, such as \cite{maleki2021mobility,bahreini2021mechanisms}. 
\begin{landscape}
\scriptsize
\begin{longtable}{cccccccccccc}
\caption{Summary of existing works considering decision engine taxonomy \label{Tab:DecisionEngine}}\\
\hline
\multirow{3}{*}{Ref}                             & \multicolumn{11}{c}{Decision Engine Characteristics}             \\* 
\cline{2-12}
& \multirow{2}{*}{\begin{tabular}[c]{@{}c@{}}Deployment\\Layer\end{tabular}} & \multicolumn{2}{c}{Admission Control} & \multirow{2}{*}{Placement Technique} & \multicolumn{4}{c}{Advanced Features}                                                              & \multicolumn{3}{c}{Implementation}                                                      \\* 
\cmidrule(lr){3-4}\cmidrule(lr){6-9}\cmidrule(lr){10-12}
&                                                                            & Queuing  & Dispatch                   &                                      & \begin{tabular}[c]{@{}c@{}}Mobility\\Support\end{tabular}  & \begin{tabular}[c]{@{}c@{}}Failure\\Recovery\end{tabular} & \begin{tabular}[c]{@{}c@{}}High\\Scalability\end{tabular} & \begin{tabular}[c]{@{}c@{}}High\\Adaptability\end{tabular} & Language     & \begin{tabular}[c]{@{}c@{}}Source\\Code\end{tabular} & Time Complexity   \endfirsthead 
\cmidrule(r){1-1}\cmidrule(r){2-2}\cmidrule(r){3-4}\cmidrule(lr){5-5}\cmidrule(r){6-9}\cmidrule(r){10-12}
\cite{asheralieva2019learning}          & Edge                                                                       & ND       & Single                     & ML, RL, DRL, (DQN)                   & \scalebox{1.1}{\ensuremath \Circle}    & \scalebox{1.1}{\ensuremath \Circle}                                                    & \scalebox{1.1}{\ensuremath \Circle}                                                    & \scalebox{1.1}{\ensuremath \CIRCLE}                                                  & ND           & \scalebox{1.1}{\ensuremath \Circle}                                               & Low (MP 2)        \\ 
\hdashline
\cite{bahreini2021vecman}               & Edge                                                                       & ND       & ND                         & Tr, H, Greedy                        & \scalebox{1.1}{\ensuremath \CIRCLE} & \scalebox{1.1}{\ensuremath \Circle}                                                    & \scalebox{1.1}{\ensuremath \Circle}                                                    & \scalebox{1.1}{\ensuremath \Circle}                                                     & ND           & \scalebox{1.1}{\ensuremath \Circle}                                               & Medium (MP 3)     \\
\end{longtable}
\end{landscape}
%%%%%%%%%%%%%%%%%%%
\subsubsection{Simulation}
Simulators keep the advantages of analytical tools while improving the credibility of evaluations by simulating the dynamics of resources, applications, and environments. In the literature, iFogSim \cite{mahmud2021ifogsim2,gupta2017ifogsim} is among the most popular simulators for Fog computing \cite{lu2020optimization,kimovski2021mobility,goudarzi2020application,tuli2020dynamic}. Besides, several researchers have used Cloudsim \cite{calheiros2011cloudsim} such as \cite{xu2019computation,deng2020optimal} or SimPy\footnote{https://simpy.readthedocs.io/en/latest/} such as \cite{nan2018dynamic,gazori2020saving} to simulate their scenarios in Fog computing. 
\subsubsection{Practical}
The most credible approach for the evaluation of proposals is practical implementation. However, due to high monetary cost, implementation time, and reproducibility time, it is not the most efficient approach for different scenarios, especially evaluations requiring high scalability. In the literature, few works such as \cite{shekhar2019urmila,gedawy2020ramos,selimi2019lightweight,badri2019energy} evaluated their proposals using small-scale practical implementations.
\subsubsection{Hybrid}
In this approach, researchers evaluate their proposals using practical implementations in small-scale and simulators or analytical tools in large-scale. Although implementation and reproducibility time of this approach is high, it provides high scalability and credibility. In the literature, few works such as \cite{goudarzi2021distributed,ma2021towards,qiu2020distributed} follow the hybrid approach.
\subsection{Metrics}
The metrics used in performance evaluation in Fog computing are directly or indirectly related to the optimization parameters and system properties. Based on the nature and popularity of these metrics in the literature, we categorize them into 1) \textbf{time} (e.g., deadline, response time, execution time, makespan) \cite{deng2020optimal,liu2021efficient,deng2021fogbus2,min2019learning}, 2) \textbf{energy} (e.g., battery percentage, saved energy) \cite{bahreini2021vecman,huang2018distributed,huang2018distributed,chen2018optimized}, 3) \textbf{monetary cost} (e.g., service cost, switching cost) \cite{asheralieva2019learning,nan2018dynamic,ouyang2018follow,mouradian2019application}, and 4) \textbf{other} metrics (e.g., number of interrupted tasks, resource utilization, throughput, deadline miss ratio) \cite{hong2019multi,goudarzi2021distributedmigration,bashir2019resource,cai2021failure}. Also, we consider 5) \textbf{decision overhead} as an important evaluation metric to study the overhead of proposals (often in terms of time and energy), used in some works such as \cite{goudarzi2021distributed,kimovski2021mobility,wu2019efficient,pan2021multi}.  
\subsection{Discussion}
In this section, we describe the lessons that we have learned regarding identified elements in the performance evaluation of the current literature. Besides, we identify several research gaps accordingly. Table~\ref{Tab:Performance Evaluation} provides a summary of characteristics related to performance evaluation in Fog computing.
\subsubsection{\textbf{Lessons learned}} Our findings regarding the performance evaluation in the surveyed works are briefly described in what follows:
\paragraph{1} 
More than half of the works used the simulation as their performance evaluation approach while 30\% of the proposals used an analytical approach. The practical and hybrid approaches equally share the rest of 20\% of the works. For the analytical approach, the most of works used Matlab or Python programming languages, while Java and Python are mostly used for the simulation approach. In practical and hybrid approaches, Java and Python are equally employed in proposals.
\paragraph{2}
As the performance evaluation metric, time and its variations (e.g., response time, makespan) are used in more than 80\% of the works. The second-highest-used metric is energy at 35\%. However, the decision overhead and cost are only studied in 15\% of the works. Besides, less than 5\% of the proposals studied the performance of their scheduling technique using all the identified metrics.  
\subsubsection{Research Gaps}
We have identified several open issues for further investigation that are discussed below:
\paragraph{1} 
Although the monetary costs of sensors and edge devices (e.g., Rpi, Jetson Platform) have reduced and they are highly available in different configurations, compared to a few years ago, the majority of proposals still consider analytical tools and simulators as their only approach for performance evaluation. While some works have considered practical and hybrid approaches for the performance evaluation of their work, further efforts are required to study the dynamics of the system, resource contention, and collaborative execution of the application in real environments, especially considering new machine learning techniques such as DRL and DDRL \cite{goudarzi2021distributed,qiu2020distributed}.
\paragraph{2} 
The decision overhead of proposals has direct effects on users and resources in terms of the startup time of requested services and resource utilization. To illustrate, not only do healthcare applications require low response time, but they also need low startup time, especially for critical applications such as emergency-related applications (e.g., heart-attack prediction and detection). Also, the overhead of proposals can severely affect the resource usage and energy consumption of servers, especially battery-constrained ones. Among the techniques considered decision overhead as a metric, they mostly focus on time while other metrics (e.g., energy, cost) need further investigation.
\section{Scheduling Technique: Important Design Options}
\label{Sec:designOptions}
In this section, we discuss the real-world characteristics of application structure and environmental architecture and accordingly present several guidelines for designing a scheduling technique.
%
%\subsection{Real-world Characteristics}
%Identifying the real-world characteristics of application structure and environmental %architecture elements lead to designing a more realistic and efficient decision %engine.
%
%\subsection{Application Structure}
\paragraph{1}
The number of IoT applications is constantly increasing in different application domains. The majority of these applications are defined as a set of dependent modules/services \cite{zhao2021offloading}. Besides, sharing and reusing modules/services for faster development and better management of applications is of paramount importance. Moreover, dependent IoT applications are usually modeled as a graph of tasks and their respective invocations. In this case, IoT applications with monolithic and independent design can also be defined as an application graph with only one module and an application graph with several modules where the size of invocations is zero, respectively. Hence, we consider IoT applications with dependent modules/services (i.e., modular and loosely-coupled categories) as the main architectural design choices in the application structure. Accordingly, the decision engine requires a component for identifying and satisfying the constraint among modules/services.
\newpage
\scriptsize
\begin{longtable}{ccccccc||ccccccc}
\caption{Summary of existing works considering performance evaluation taxonomy \label{Tab:Performance Evaluation}}\\ 
\hline
\multirow{4}{*}{Ref}                    & \multicolumn{6}{c||}{Performance Evaluation}                            &  \multirow{4}{*}{Ref}    & \multicolumn{6}{c}{Performance Evaluation}                                                                                                                                                                 \\*
\cline{2-7}\cline{9-14}
& \multicolumn{6}{c||}{Metrics} & & \multicolumn{6}{c}{Metrics} \\*
\cline{3-7}\cline{10-14}
& \multirow{1}{*}{Approach}                                                             & DO      & T       & E       & C       & O                                                                  &                                                  & \multirow{1}{*}{Approach}                                                             & DO      & T       & E       & C       & O                                                                                           \endfirsthead 
\hline
\cite{asheralieva2019learning} & \begin{tabular}[c]{@{}c@{}}Sim\\(OPNET)\end{tabular}                 & \scalebox{1.1}{\ensuremath \Circle}  & \scalebox{1.1}{\ensuremath \Circle}  & \scalebox{1.1}{\ensuremath \Circle}  & \scalebox{1.1}{\ensuremath \CIRCLE}   & \scalebox{1.1}{\ensuremath \Circle}                                                             & \cite{mahmud2019context}                & \begin{tabular}[c]{@{}c@{}}Hybrid\\(Sim, Prac)\\(iFogSim)\end{tabular} & \scalebox{1.1}{\ensuremath \Circle}  & \scalebox{1.1}{\ensuremath \CIRCLE}   & \scalebox{1.1}{\ensuremath \Circle}  & \scalebox{1.1}{\ensuremath \Circle}  & \begin{tabular}[c]{@{}c@{}}Resource\\Overhead\end{tabular}                                  \\ 
\hdashline
\cite{bahreini2021vecman}      & ND                                                                   & \scalebox{1.1}{\ensuremath \Circle}  & \scalebox{1.1}{\ensuremath \Circle}  & \scalebox{1.1}{\ensuremath \CIRCLE}   & \scalebox{1.1}{\ensuremath \Circle}  & \scalebox{1.1}{\ensuremath \Circle}                                                             & \cite{el2018computational}              & Analytic                                                             & \scalebox{1.1}{\ensuremath \Circle}  & \scalebox{1.1}{\ensuremath \Circle}  & \scalebox{1.1}{\ensuremath \Circle}  & \scalebox{1.1}{\ensuremath \Circle}  & \begin{tabular}[c]{@{}c@{}}Weighted\\Cost\end{tabular}                                      \\
\end{longtable}
\normalsize
\paragraph{2}
Besides, in a real-world scenario, application modules have different characteristics (e.g., computation size, input size, ram usage). Thus, the best assumption for application modules is applications with heterogeneous granularity specifications. As the number of contributing parameters and the dynamicity of the application elements increases, capturing the application parameters with temporal patterns for efficient scheduling decisions becomes more complex \cite{tuli2020dynamic,goudarzi2021distributed}. Although traditional-based placement techniques (e.g., heuristic, meta-heuristic) often work well in general scenarios, they fail to adapt to continuous changes and dynamic contexts. ML-based decision engines, such as RL, can more efficiently work in a dynamic context and provide higher adaptability.
\paragraph{3}
In large-scale Fog computing environments, numerous IoT applications with different workload models and hybrid CCR may exist. Hence, the decision engine requires an admission control component with an appropriate queuing mechanism (based on application requirements) to manage diverse incoming requests and prioritize them for making the decision.
\paragraph{4}
Regarding the environmental architecture, the most generalized scenario is when the environment consists of several heterogeneous IoT devices, several heterogeneous FSs, and multiple heterogeneous CSs. Also, the required mechanisms for intra-tier and inter-tier cooperation among servers should be embedded to support diverse IoT application scenarios, such as mobility. Besides, multiple distributed servers can collaboratively provide better performance for the execution of IoT applications. Moreover, different fault domains can be prepared to improve the availability of services. However, as the number of IoT applications and available servers in the environment increase, the complexity of making decisions increases. Hence, the optimal scheduling decision cannot be obtained in a timely manner. Consequently, other placement techniques such as heuristics and ML-based techniques should be employed to obtain the scheduling decision in a reasonable time.
\paragraph{5}
The decision engine can be implemented as a set of distributed services/microservices. A decision engine developed as a monolithic application may not be able to be deployed on a single server, especially on resource-limited FSs. Hence, distributed deployment of decision engine components on several distributed servers can provide several benefits: 1) more efficient deployment of resource-limited devices, 2) provides better fault tolerance 3) offers better scalability 4) support different deployment models (e.g., deployment of decision engine on FSs, CSs, or hybrid on both FSs and CSs). Hybrid deployment of decision engine components on both FSs and CSs can lead to a better user experience for end-users. To illustrate, applications requiring low latency and startup time can be managed at the low-level FSs (i.e., at the Edge), and then be scheduled based on the decision engine deployed at the Edge. However, application requests that are insensitive to latency or startup time can be forwarded to CSs for scheduling. 
\paragraph{6}
Regardless of application and environmental characteristics, failure recovery mechanisms and policies should be integrated into any decision engine. Independent failures and the non-deterministic nature of any components (either hardware or software) in distributed systems cause the most impactful issues in distributed systems. If the decision engine, which manages the scheduling and execution of incoming IoT application requests, does not have an appropriate failure recovery mechanism, the smooth execution of the whole system stalls. 
%
%\subsection{Environmental Architecture}
%

%\subsection{Design Guidelines}
%
\section{Future Research Directions}
\label{Sec:futureDirections}
This section presents future research directions, guiding researchers to further progress in the field of Fog computing.
\paragraph{\textbf{Microservices-based applications}}
The popularity of microservices for the deployment of IoT applications is due to their loosely-coupled design, modularity, and the capability of microservices to be shared among multiple IoT applications. But, it may incur data consistency and data privacy challenges. To overcome these challenges, the placement techniques should consider the context of applications and data before sharing microservices. 
\paragraph{\textbf{Practical Container orchestration in Fog computing}}
Orchestrating container-based IoT applications is well studied in the cloud computing paradigm. However, in Fog computing, in which CSs and FSs collaborate to run an application, several deployment models of orchestration techniques are available. To illustrate, the master node can either be deployed on a FS or CS. When the master node runs on a FS, the communication overhead and latency for end-users will be reduced. However, the master node will use the most of resources on the FS for the cluster management, especially for resource-limited FSs. Also, when the master runs on a CS, the startup time and application latency will be negatively affected. Thus, based on the application structure and its goal, different container orchestration models should be studied to find the best deployment model according the application scenario. Several practical studies can be conducted to find which deployment model is suitable for each IoT application scenario in terms of communication overhead, the startup time of services, memory footprint, failure management, load balancing, and scheduling.
\paragraph{\textbf{Hybrid scheduling decision engines}} Usually, decision engines only use one placement technique for different IoT applications. However, the requirements of IoT applications are heterogeneous, where one application is sensitive to startup time while the extremely high accuracy is not important, or vice versa. Besides, decision engines should be adapted to work with either single or batch placement approaches. Hence, context-aware decision engines with a suite of placement techniques can be implemented to address the requirements of different IoT applications.  
\paragraph{\textbf{Systems for ML}}
Due to advancements in ML techniques and their rapid adoptions across many IoT applications, it creates new demand for specialized hardware resources and software frameworks (e.g., Nvidia GPU-powered Jetson, Google Coral Edge Tensor Processing Unit (Edge TPU)) for Fog computing. New systems and software frameworks should be built to support the massive computational requirement of these AI workloads. Besides, these systems can be a potential target for the deployment of decision engines due to their high computational capacity.
\paragraph{\textbf{ML for systems}}
While ML systems themselves are becoming mature and adopted into many critical application domains, it is equally important to use these ML techniques to design and operate large-scale systems. Adopting the ML techniques to solve different resource management problems in Edge/Fog and Cloud is crucial to managing these complex infrastructures and workloads. Moreover, majority of ML techniques are not optimized to run on resource-constrained devices. To illustrate, consider an efficient ML model trained for resource management. Many resource-constrained devices require full integer quantization to run the trained model. However, post quantization of trained models is not always possible and in some cases they cannot be efficiently converted. As a result, a study on requirements for the efficient execution of resource management ML models on resource-limited FSs should be conducted.
\paragraph{\textbf{Thermal management}}
The temperature of FSs (e.g., racks of Rpi or Nvidia Jetson platform), especially those executing large workloads, increases significantly. So, the cooling systems should be embedded to avoid system breakdown. Hence, a study on the temperature of these devices based on their main processing and communication modules can be conducted to find the respective temperature dynamics in different application scenarios and workloads. Moreover, lightweight thermal management software systems for FSs can be designed to control the temperature dynamics of devices. Also, the thermal index can be added as an important optimization/decision parameter alongside other currently available parameters (e.g., time, energy, cost) for the placement techniques.
\paragraph{\textbf{Trade-off between execution cost of IoT devices and resource providers}}
The goal of scheduling algorithms is to minimize the execution cost of applications either from IoT or resource providers' perspectives. However, some parameters such as energy consumption or carbon footprint should be considered from both perspectives. Hence, not only is minimizing these parameters from either perspective critical to reducing total energy consumption, but a trade-off parameter between the execution cost of IoT devices and resource providers can be designed, aiming at total energy or carbon footprint minimization.  
\paragraph{\textbf{Privacy aware and adaptive decision engines}}
Data-driven and distributed scheduling approaches are gaining popularity due to their high adaptability and scalability. However, sharing raw data of users or systems incurs privacy issues. To illustrate, in DDRL-based scheduling techniques, sharing experiences of multiple agents significantly reduce the exploration costs and improve convergence time of DDRL agents while incurring privacy concerns when raw agents' experiences are shared. Accordingly, privacy-aware mechanisms for sharing such data (e.g., agents' experiences) can be integrated with these highly adaptive distributed scheduling techniques.
\section{Summary}
\label{Sec:conclusion}
In this paper, we mainly focused on scheduling IoT applications in Fog computing environments. We identified several main perspectives that play an important roles in scheduling IoT applications, namely application structure, environmental architecture, optimization characteristics, decision engines properties, and performance evaluation. Next, we separately identified and discussed the main elements of each perspective and provided a taxonomy and research gaps in the recent literature. Finally, we highlighted several future research directions for further improvement of Fog computing.         
%\begin{acks}
%To Robert, for the bagels and explaining CMYK and color spaces.
%\end{acks}

\bibliographystyle{ACM-Reference-Format}
\bibliography{sample-base}

\end{document}